\documentclass[pra,twocolumn,aps,floatfix,showpacs,tightenlines,showpacs,superscriptaddress]{revtex4}

\usepackage{amssymb,amsmath,amsthm,amsbsy,epsfig,color,graphicx,times}

\usepackage{stackengine}

\begin{document}

\title{Entanglement and quantum correlations in many-body systems: \\
       a unified approach via local unitary operations}

\author{M. Cianciaruso}
\affiliation{Dipartimento di Fisica ``E. R. Caianiello'', Universit\`a degli Studi di Salerno,
Via Giovanni Paolo II 132, I-84084 Fisciano (SA), Italy}
\affiliation{INFN Sezione di Napoli, Gruppo collegato di Salerno, Italy}

\author{S. M. Giampaolo}
\affiliation{International Institute of Physics, UFRN, Av. Odilon Gomes de Lima 1722, 59078-400 Natal, Brazil}
\affiliation{Dipartimento di Ingegneria Industriale, Universit\`a degli Studi di Salerno, Via Giovanni Paolo II 132, I-84084 Fisciano (SA), Italy}

\author{W. Roga}
\affiliation{Department of Physics, University of Strathclyde, John Anderson Building, 107 Rottenrow, Glasgow G4 0NG, United Kingdom}
\affiliation{Dipartimento di Ingegneria Industriale, Universit\`a degli Studi di Salerno, Via Giovanni Paolo II 132, I-84084 Fisciano (SA), Italy}

\author{G. Zonzo}
\affiliation{Dipartimento di Fisica ``E. R. Caianiello'', Universit\`a degli Studi di Salerno, Via Giovanni Paolo II 132, I-84084 Fisciano (SA), Italy}

\author{M. Blasone}
\affiliation{Dipartimento di Fisica ``E. R. Caianiello'', Universit\`a degli Studi di Salerno,
Via Giovanni Paolo II 132, I-84084 Fisciano (SA), Italy}
\affiliation{INFN Sezione di Napoli, Gruppo collegato di Salerno, Italy}

\author{F. Illuminati}
\thanks{Corresponding author: illuminati@sa.infn.it}
\affiliation{Dipartimento di Ingegneria Industriale, Universit\`a degli Studi di Salerno, Via Giovanni Paolo II 132, I-84084 Fisciano (SA), Italy}
\affiliation{INFN Sezione di Napoli, Gruppo collegato di Salerno, Italy}

\date{October 22, 2015}

\begin{abstract}
Local unitary operations allow for a unifying approach to the quantification of quantum correlations among the constituents of a bipartite quantum
system. For pure states, the distance between a given state and its image under least-perturbing local unitary operations is a {\em bona fide} measure of quantum entanglement, the so-called entanglement of response, which can be extended to mixed states via the convex roof construction. On the other hand, when defined directly on mixed states perturbed by local unitary operations, such a distance turns out to be a {\em bona fide} measure of quantum correlations, the so-called discord of response. Exploiting this unified framework, we perform a detailed comparison between two-body entanglement and two-body quantum discord in infinite $XY$ quantum spin chains both in symmetry-preserving and symmetry-breaking ground states as well as in thermal states at finite temperature. The results of the investigation show that in symmetry-preserving ground states the two-point quantum discord dominates over the two-point entanglement, while in symmetry-breaking ground states the two-point quantum discord is strongly suppressed and the two-point entanglement is essentially unchanged. In thermal states, for certain regimes of Hamiltonian parameters, we show that the pairwise quantum discord and the pairwise entanglement can increase with increasing thermal fluctuations.
\end{abstract}


\maketitle

\section{Introduction}

Quantum correlations arise from the combination of the superposition principle and the tensor product structure of the Hilbert space associated
with a composite quantum system. For pure states, they are entirely captured by entanglement. On the other hand, in the case of mixed states, the situation becomes more involved, as there can exist mixed separable (i.e. non-entangled) states that nevertheless can display non classical features~\cite{OZ2001,HV2001,PHH2008}. The existence of such states suggests that the total amount of quantum correlations is not, in general, quantified only by the entanglement but needs to be characterized also in terms of another quantity, related to quantum state distinguishability, the so-called quantum discord.

Entanglement and discord are fundamental resources for quantum information and quantum metrology~\cite{NC2000,HHHH2009,LBAW2008,Girolami2014}, as well as quite useful tools for the characterization of quantum phases in many-body systems~\cite{AFOV2008,WRR2012,SdOA2013}. For instance, topologically ordered phases cannot be characterized by the Landau-Ginzburg paradigm based on symmetry breaking and local order parameters, but rather by the long-range entanglement properties featured by the ground state of the system~\cite{Balents2012,Kitaev2006,Levin2006,GH2014}. Within this generalized framework, quantum correlations (entanglement and discord) in many-body ground states allow for the most fundamental characterization of complex quantum systems. In fact, even for systems that do not feature exotic phases and nonlocal quantum orders, the investigation of ground-state patterns of entanglement and discord can provide a deeper understanding of locally ordered phases associated to spontaneous symmetry breaking~\cite{VLRK2003,LRV2004,GAI2008,GAI2009,GAI2010,GMASI2013,GH2013,HOG2013,TRHA2011,CRLB2013,CFGI2014,HGI2015}.


In spite of the ongoing efforts to characterize quantum ground states by analyzing their quantum correlations, a systematic comparative study of the behavior of entanglement and discord in quantum many-body systems is still lacking. In the present work we carry out such direct comparison within the powerful unified framework to the quantification of entanglement and quantum correlations, based on the formalism of local unitary operations, introduced in Refs.~\cite{GI2007,MAGGDI2011,GHARIB2012,GSRBI2013,RGI2013}.
In the above-mentioned works, it has been shown that the distance between a given state and the state obtained from it by applying a least perturbing local unitary operation is a {\em bona fide} measure of entanglement, the so-called entanglement of response (or {\em unitary entanglement})~\cite{MAGGDI2011,GI2007,GHARIB2012}, in the case of pure states. In the case of mixed states, it is a {\em bona fide} measure of quantum correlations (discord), the so-called discord of response~\cite{RGI2013,GSRBI2013}.

We apply these distance-based measures to investigate the ground-state behavior of pairwise entanglement and discord in the one-dimensional $XY$ models in a transverse magnetic field with periodic boundary conditions. Within the set of well-behaved distances, featuring the correct properties of monotonicity under completely positive and trace preserving (CPTP) dynamical maps, we pick the trace distance (other possible relevant choices would include, e.g., the relative entropy, the Bures, and the Hellinger distance). We show that the pairwise discord of response is always larger than the pairwise entanglement of response, and strongly dominates it in symmetry-preserving ground states, particularly for large inter-particle distances and at the factorizing field (this last fact being trivial, since at the factorizing field the pairwise entanglement always vanishes identically).

For symmetry-breaking ground states, we observe that while the hierarchy between discord and entanglement continues to hold, nevertheless, compared to the symmetry-preserving case, the pairwise discord is strongly suppressed while the pairwise entanglement remains either unchanged or increases slightly for decreasing values of the external field below the factorization point. Moving from the trace distance, which is monotonically non-increasing under CPTP  maps, to the Hilbert-Schmidt distance, that does not share such a property, we show that in symmetry-breaking ground states the physically correct hierarchy is reversed: the pairwise discord of response is dominated by the entanglement of response. This unphysical result thus provides an important illustration of the fact that the Hilbert-Schmidt metric does not yield a proper and correct quantification of quantum correlations (a first well-known example was provided earlier by Piani in Ref.~\cite{P2012}).

The paper is organized as follows. In Section~\ref{sec:definitionsandnotationsofstellarentandquant} we review the unifying approach to
the quantification of quantum correlations based on local unitaries, by recalling the definitions of the entanglement and discord of response. In Section~\ref{sec:XYmodel} we recall the main features of the one-dimensional $XY$ models in transverse field with periodic boundary
conditions. In Sections~\ref{sec:symmetricgroundstate}, \ref{sec:symmmetrybrokengroundstate} and \ref{sec:thermalstate} we perform the comparison
between the entanglement of response and the discord of response for spin pairs in infinite $XY$ chains (thermodynamic limit), respectively in symmetry-preserving and symmetry-breaking ground states, as well as in thermal states at finite temperature. Conclusions and outlook are discussed in Section~\ref{sec:conclusions}.

\section{Entanglement and discord of response}\label{sec:definitionsandnotationsofstellarentandquant}

We start by briefly reviewing some basic definitions and results concerning the quantification of entanglement and quantum correlations via local unitary operations. Throughout the present work, we will focus on a bipartite quantum system $AB$ composed of two distinguishable subsystems $A$ and $B$. Such quantum system is associated with an Hilbert space \mbox{$\mathcal{H}=\mathcal{H}_A\otimes\mathcal{H}_B$}
which is the tensor product of the Hilbert spaces pertaining to each subsystem, so that
\mbox{$d\equiv\dim\mathcal{H}=d_A d_B$}. Moreover, the space of
states of $AB$ is characterized by the convex set of density operators (i.e. semi-positive definite and trace-class operators with unit trace) on $\mathcal{H}$, whose extremal points are the unit-trace projectors over $\mathcal{H}$ that represent pure states.


Let us denote by $\rho_\Phi^{AB}\equiv|\Phi^{AB}\rangle\langle\Phi^{AB}|$ and $\Lambda$, respectively, a generic pure state of the bipartite
quantum system $AB$ and the set of local unitary operators $U_A \equiv U_A \otimes \mathbb{I}_B$ such that $\mathbb{I}_B$ is the identity operator
on $\mathcal{H}_B$ and $U_A$ is any unitary operator on $\mathcal{H}_A$ whose spectrum is given by the $d_A$-th complex roots of unity. The
{\em entanglement of response}~\cite{GI2007,MAGGDI2011} of $\rho_\Phi^{AB}$,
$E\left(|\Phi^{AB}\rangle \right) $, is defined by:
\begin{equation}
E\left(|\Phi^{AB}\rangle \right) \equiv  \min_{U_A \in \Lambda}  D_{Tr}\left(\rho_\Phi^{AB},\tilde{\rho}_\Phi^{AB} \right)^2 \; ,
\end{equation}
where $\tilde{\rho}_\Phi^{AB}\equiv U_A\rho_\Phi^{AB} U_A^\dagger$ and $D_{Tr}(\rho,\sigma)\equiv\frac{1}{2} Tr|\rho-\sigma|$ is the trace distance between the states $\rho$ and $\sigma$.
In other words, when the whole quantum system $AB$ is in a pure state $\rho_\Phi^{AB}$, the entanglement of response quantifies the quantum correlations
between parts $A$ and $B$ in terms of the distinguishability between the state $\rho_\Phi^{AB}$ and the state $\tilde{\rho}_\Phi^{AB}$ obtained from $\rho_\Phi^{AB}$ by applying to it a minimally perturbing local unitary transformation.


There are at least two distinct ways to extend the entanglement of response to mixed states: the convex roof extension, which then identifies the entanglement of response of mixed states, and the {\em discord of response} defined directly as the distance between a given mixed state and the one obtained from it through the action of the least perturbing local unitary operation~\cite{RGI2013,GSRBI2013}. More precisely, the {\em entanglement of response} of a bipartite mixed state $\rho^{AB}$, $E\left(\rho^{AB} \right) $, is defined as:
\begin{equation}\label{eq:stellarentanglementconvexroofextention}
E\left(\rho^{AB} \right) \equiv \min_{\left\lbrace |\Phi_i^{AB}\rangle, p_i\right\rbrace } \sum_i p_i E\left(|\Phi_i^{AB}\rangle \right) \; ,
\end{equation}
where the minimization is performed over all the decompositions of $\rho^{AB}$ in pure states $\sum_i p_i |\Phi_i^{AB}\rangle\langle
\Phi_i^{AB}| = \rho^{AB}$, $p_i\geq 0$, $\sum_i p_i = 1$.
On the other hand, the {\em discord of response} of a bipartite state $\rho^{AB}$, $Q(\rho^{AB})$, is defined as~\cite{RGI2013} (see also Ref.~\cite{GSRBI2013} for earlier related work):
\begin{equation}
Q(\rho^{AB}) \equiv  \min_{U_A\in \Lambda} D_{Tr}\left(\rho^{AB},\tilde{\rho}^{AB} \right)^2 ,
\end{equation}
where, as in the case of pure states, $\tilde{\rho}^{AB}\equiv U_A\rho^{AB} U_A^\dagger$.

Therefore, the entanglement and the discord of response quantify different aspects of bipartite quantum correlations via two different uses of local unitary operations. The discord of response arises by applying local unitaries directly to the generally mixed state $\rho^{AB}$, while the entanglement of response stems from the application of local unitaries to pure states. By virtue of their common origin, it is thus possible to perform a direct comparison between these two quantities.

In terms of the trace distance, the two-qubit entanglement of response is simply given by the squared concurrence~\cite{W1998,RGI2013}, whereas the two-qubit discord of response relates nicely to the geometric discord~\cite{NPA2013}, whose closed formula is known only for a particular class of two-qubit states~\cite{CTG2013}, although it can be computed for a more general class of two-qubit states through a very efficient numerical optimization.

\section{$XY$ MODELS}\label{sec:XYmodel}

In this section we recall some key aspects of the quantum many body
systems we shall focus on, that is the one-dimensional anti-ferromagnetic $XY$ models in transverse field with periodic boundary
conditions~\cite{LSM1961,P1970,BMD1970,BM1971,JM1971}. Such quantum spin models
consist of a periodic chain of $N$ $\frac{1}{2}$-spins, with anisotropic nearest-neighbor spin-spin interactions competing with a transverse magnetic field, whose dynamics is governed by the following Hamiltonian:
\begin{equation}\label{eq:XYmodelhamiltonian}
H\! =\!\sum_{i=1}^{N}\! \left[\left(\frac{1+\gamma }{2}\right) \sigma_i^x \sigma_{i+1}^x \!+\! \left(\frac{1-\gamma }{2}\right)\!\sigma_i^y
\sigma_{i+1}^y \!-\! h \sigma_i^z\right]\!.
\end{equation}
Here $\sigma_i^\alpha$, $\alpha = x, y, z$, are the Pauli matrices on site $i$, $\gamma$ is the anisotropy in the $xy$ plane, $h$ is the
strength of the transverse magnetic field, and the periodic boundary conditions imply that $\sigma_{N+1}^\alpha\equiv \sigma_1^\alpha$.
The $XY$ models reduce to the isotropic $XX$ model and to the Ising model for $\gamma = 0$ and $\gamma = 1$, respectively.

Regardless of the value of $\gamma$,
in the thermodynamic limit, these models feature a quantum phase transition at $h=h_c=1$. For $h>h_c=1$ and for any value of $\gamma$, the ground state space is non-degenerate and there is a finite gap in the energy spectrum between the ground state and the first excited state. On the other hand,
for $h<h_c$, two different cases arise: for $\gamma=0$ the ground state space remains non-degenerate while the energy spectrum becomes gapless, whereas for $\gamma>0$ the
ground state space becomes two-fold degenerate, the energy spectrum is gapped, and the system can be characterized by a non vanishing local order
parameter $m_x=(-1)^i\langle \sigma_i^x \rangle$ (spontaneous on-site magnetization).
Besides the quantum critical point, there exists another relevant value of the external magnetic field, that is $h_f=\sqrt{1-\gamma^2}$,
the {\em factorizing field}. Indeed, at this value of $h$, the system admits a two-fold degenerate, completely factorized ground
state~\cite{KTM1982,RVFHT2005,GAI2008,GAI2009,GAI2010}. The two degenerate factorized states collapse onto a single state for $\gamma=0$. This corresponds t the isotropic, gapless $XX$ model, for which the factorizing field and the critical field coincide: $h_f = h_c = 1$.

Since our goal is to compare the two-spin entanglement of response and the two-spin discord of response, we need to determine
the pairwise reduced density matrix $\rho_{ij}$ both in symmetry-preserving and symmetry-breaking ground states, as well as thermal states at finite temperature. The pairwise reduced density matrix $\rho_{ij}$ is defined as the partial trace on the state of the whole chain with respect to all spins except those at sites $i$ and $j$. While the ground state of the entire chain is a pure state, the reduced state of a pair of spins is in general mixed. The two-site density matrix can be expanded as follows~\cite{ON2002}:
\begin{equation}
 \label{genericrho}
 \rho_{ij}=\frac{1}{4} \sum_{\alpha,\beta=0}^{3} \langle \sigma_i^\alpha \sigma_j^\beta \rangle \sigma_i^\alpha \sigma_j^\beta
\end{equation}
where $\sigma_i^0=\mathbb{I}_i$ is the identity on site $i$, and $ \langle \sigma_i^\alpha \sigma_j^\beta \rangle$ denotes the two-body correlation function between $\sigma_i^\alpha$ and $ \sigma_j^\beta $. The operator expansion in eq.~(\ref{genericrho}) depends on $15$ different correlation functions.
However, this number can be reduced by resorting to the symmetries of the Hamiltonian. Translational invariance of the lattice implies that the reduced
density matrix depends only on the inter-spin distance $r=|i-j|$. Also, since the Hamiltonian is real, $\rho_{ij}=\rho^*_{ij}$.
Finally, except for the symmetry-breaking ground states, the global phase-flip symmetry implies that  $[\sigma_i^z\sigma_j^z, \rho_{ij}]=0$.
Therefore, for both the symmetry-preserving ground states and the thermal states, the only correlation functions different from zero are $\langle \sigma_i^z \rangle$ and  $ \langle \sigma_i^\alpha \sigma_j^\alpha \rangle$ for $\alpha=x,y,z$.
Such correlation functions are readily obtained by generalizing the approach of Ref.~\cite{LSM1961} at non vanishing external
field for the finite size system, or directly from Refs.~\cite{BMD1970,BM1971} in the thermodynamic limit.

On the other hand, when dealing with the symmetry-breaking ground states, i.e. when the system is in the
thermodynamic limit, the external field is below the quantum critical point, and $\gamma>0$, the set of nonvanishing correlation functions includes also $\langle \sigma_i^x \rangle$ and $\langle \sigma_i^x \sigma_j^z \rangle $. The explicit expression of the
former was first derived in Ref.~\cite{BM1971} while $\langle \sigma_i^x \sigma_j^z \rangle $ can be evaluated by a simple generalization of the same procedure.

\section{Symmetry-preserving ground states}\label{sec:symmetricgroundstate}

\begin{figure}[t]
\includegraphics[width=8cm]{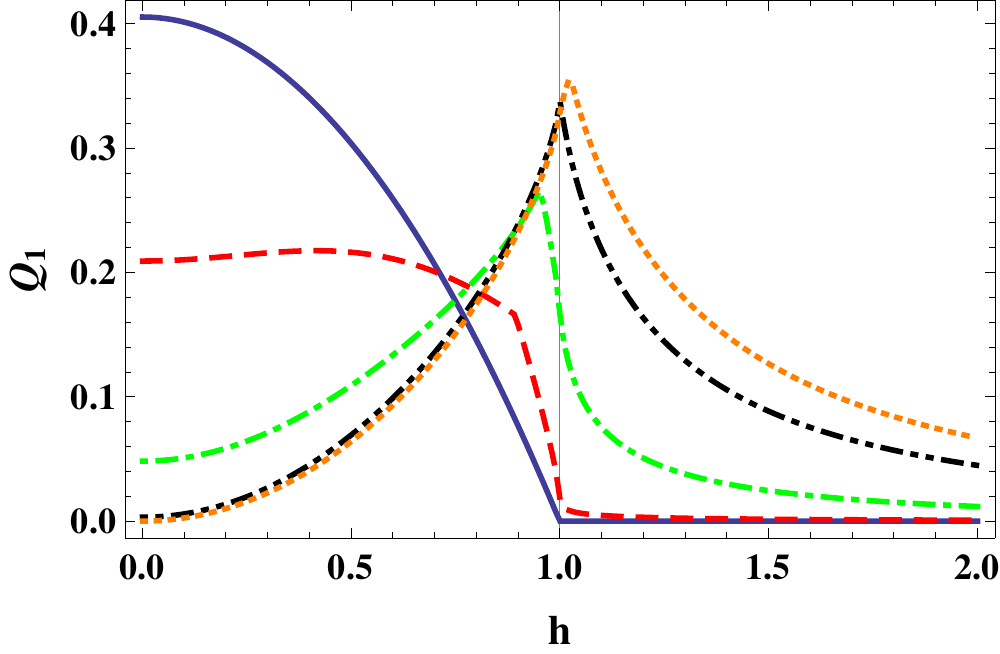}
\includegraphics[width=8cm]{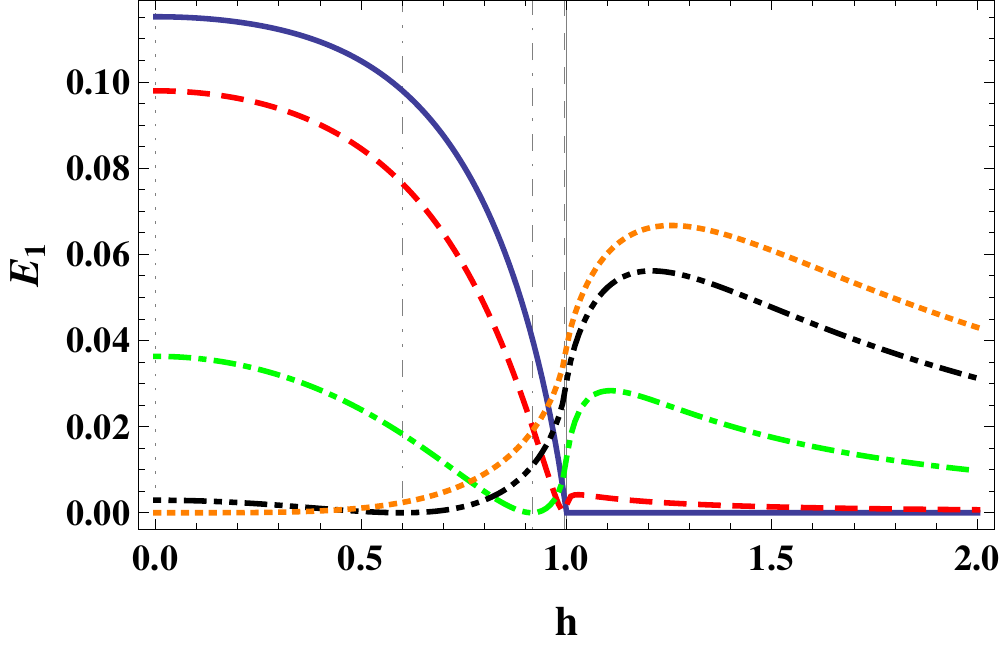}
\caption{Nearest-neighbor discord of response (upper panel) and nearest-neighbor entanglement of response (lower panel) for symmetry-preserving ground states, in the thermodynamic limit, as functions of the external field $h$, and for different values of the anisotropy $\gamma$. Solid blue curve: $\gamma=0$; dashed red curve: $\gamma=0.1$; dot-dashed green curve: $\gamma=0.4$; double-dot-dashed black curve: $\gamma=0.8$; dotted orange curve: $\gamma=1$. In the lower panel, to each of these curves, there corresponds a vertical line denoting the associated factorizing field $h_f$. In the upper panel, the solid vertical line denotes the critical field $h_c = 1$.}
\label{fig:symnearestquantandent}
\end{figure}

We first compare the two-body entanglement of response and the two-body discord of response in symmetry-preserving ground states.
For two neighboring spins, these two quantities are plotted in Fig.~\ref{fig:symnearestquantandent} as functions of the external field $h$ and for different values of the anisotropy $\gamma$. For any intermediate value of $\gamma$, the nearest-neighbor entanglement of response $E_1$ exhibits the following behavior. If $h<h_f$, $E_1$ decreases until it vanishes at the factorizing field $h=h_f$. Otherwise, if $h>h_f$, $E_1$ first increases until it reaches a maximum at some value of $h$ higher than the critical point $h_c=1$, then it decreases again until it vanishes asymptotically for very large values of $h$ in the paramagnetic phase(saturation). Overall, $E_1$ features two maxima at $h=0$ and $h>h_c$ and two minima at $h=h_f$ (factorization) and $h\rightarrow\infty$ (saturation). In the Ising model ($\gamma=1$) the point $h=0$ corresponds to a minimum,
since it coincides with the factorizing field $h_f=\sqrt{1-\gamma^2}$, while in the isotropic $XX$ model ($\gamma=0$) there is no second maximum for large fields $h>h_c$, since the ground state is always completely factorized as soon as $h \ge h_c$.

On the other hand, regardless of the value of $\gamma$, the nearest-neighbor discord of response $Q_1$ always features a single maximum. Depending on the value of $\gamma$ such maximum can be either in the ordered phase $h<h_c$ or in the disordered (paramagnetic) phase $h>h_c$, moving towards higher values of $h$ with increasing $\gamma$. Remarkably, $Q_1$ never vanishes at the factorizing field, except in the two extreme cases of $\gamma=0,1$. Indeed, at the factorizing field $h=h_f$ and for any $\gamma\neq 0,1$, the symmetry-preserving ground state is not completely factorized but rather
is a coherent superposition of the two completely factorized symmetry-breaking ground states. Consequently, while the two-body entanglement of response must vanish in accordance with the convex roof extension, the two-body discord of response remains always finite.

\begin{figure}[t]
\def\upperinsetquant{\includegraphics[height=2cm]{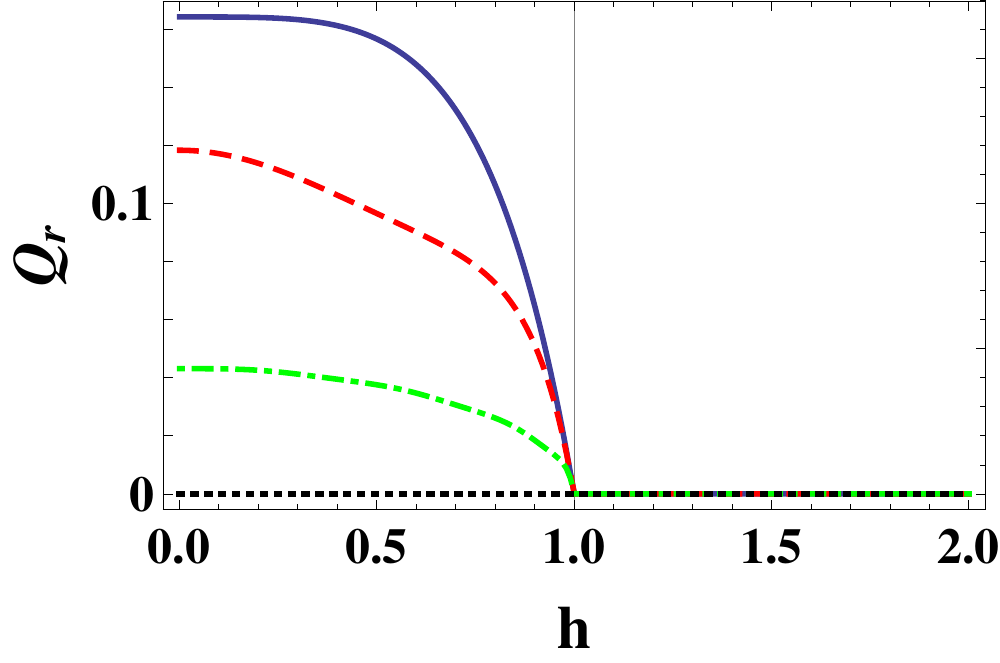}}
\stackinset{r}{8pt}{t}{7pt}{\upperinsetquant}{\includegraphics[width=8cm]{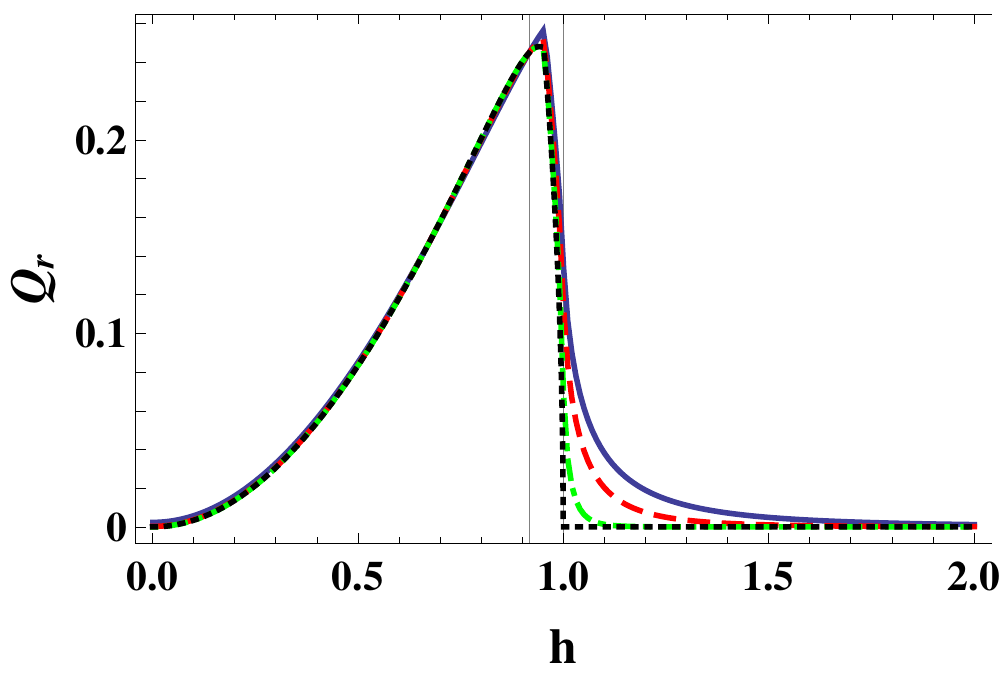}}
\def\upperinsetent{\includegraphics[height=2cm]{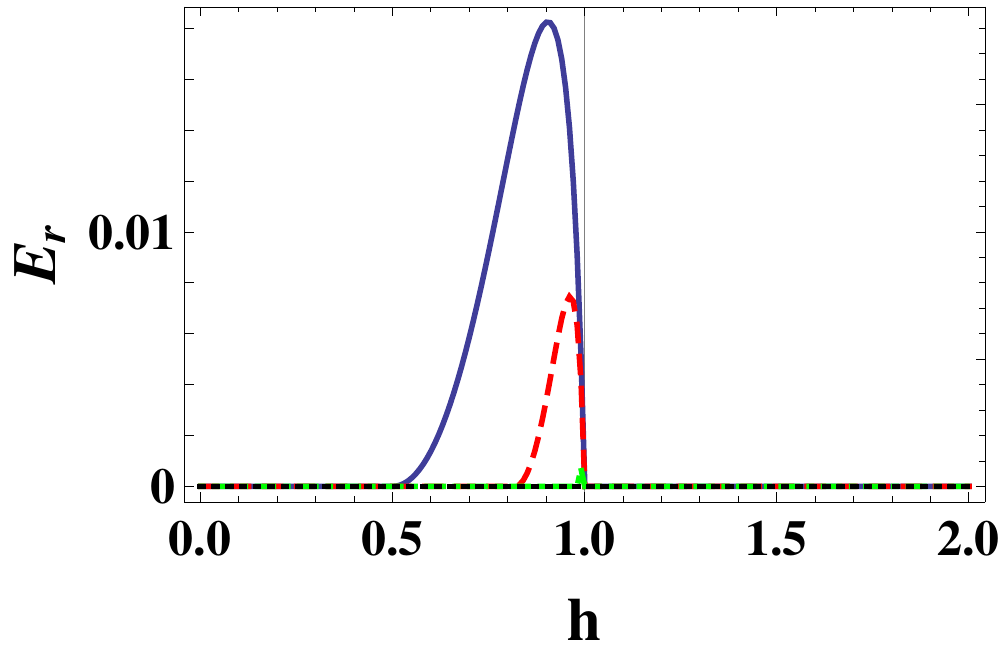}}
\stackinset{r}{8pt}{t}{7pt}{\upperinsetent}{\includegraphics[width=8cm]{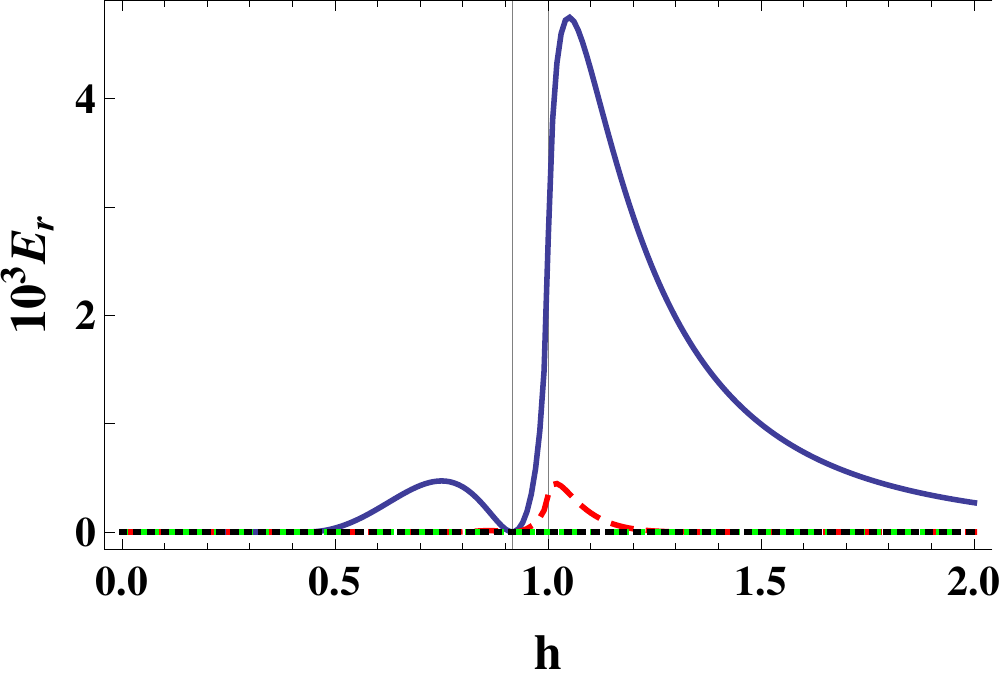}}
\caption{Two-body discord of response (upper panel) and two-body entanglement of response (lower panel) for symmetry-preserving ground states, in the thermodynamic limit, as functions of the external field $h$, in the case of $\gamma=0.4$, for different inter-spin distances $r$. Solid blue curve: $r=2$; dashed red curve: $r=3$; dot-dashed green curve: $r=8$; dotted black curve: $r=\infty$. In both panels, the two solid vertical lines correspond, respectively, to the factorizing field (left) and to the critical field (right). Inset (both panels): same, but with $\gamma=0$; the solid vertical line corresponds to the critical point.}
\label{fig:symnquantandentversushatvariousr}
\end{figure}

\begin{figure}[t]
\def\upperinsetquant{\includegraphics[height=2cm]{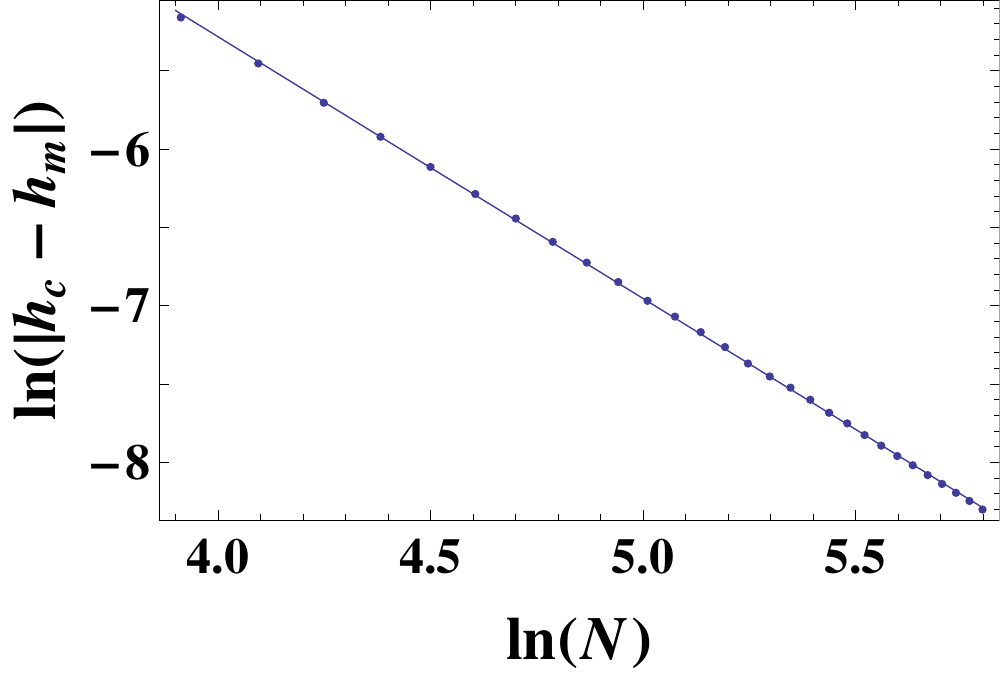}}
\def\lowerinsetquant{\includegraphics[height=2cm]{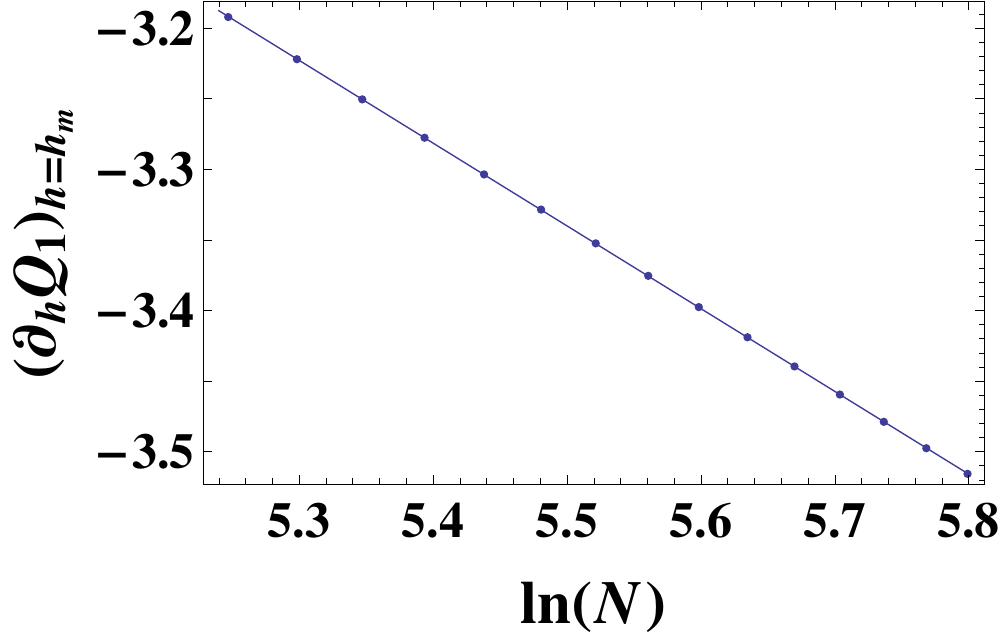}}
\stackinset{r}{8pt}{t}{7pt}{\upperinsetquant}{%
\stackinset{r}{8pt}{b}{35pt}{\lowerinsetquant}{%
\includegraphics[height=5.5cm]{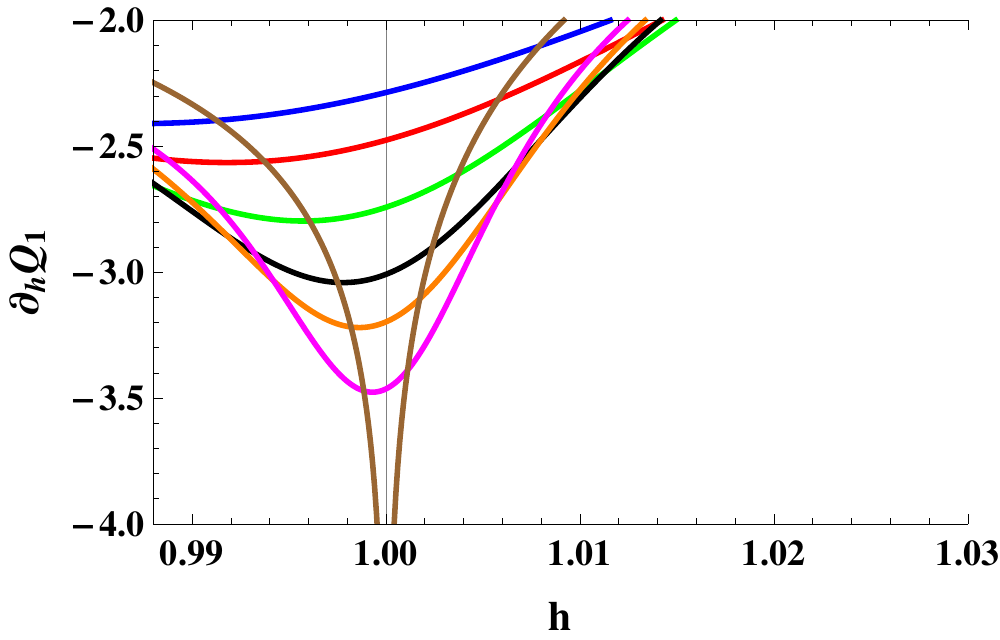}%
}}
\def\upperinsetent{\includegraphics[height=2cm]{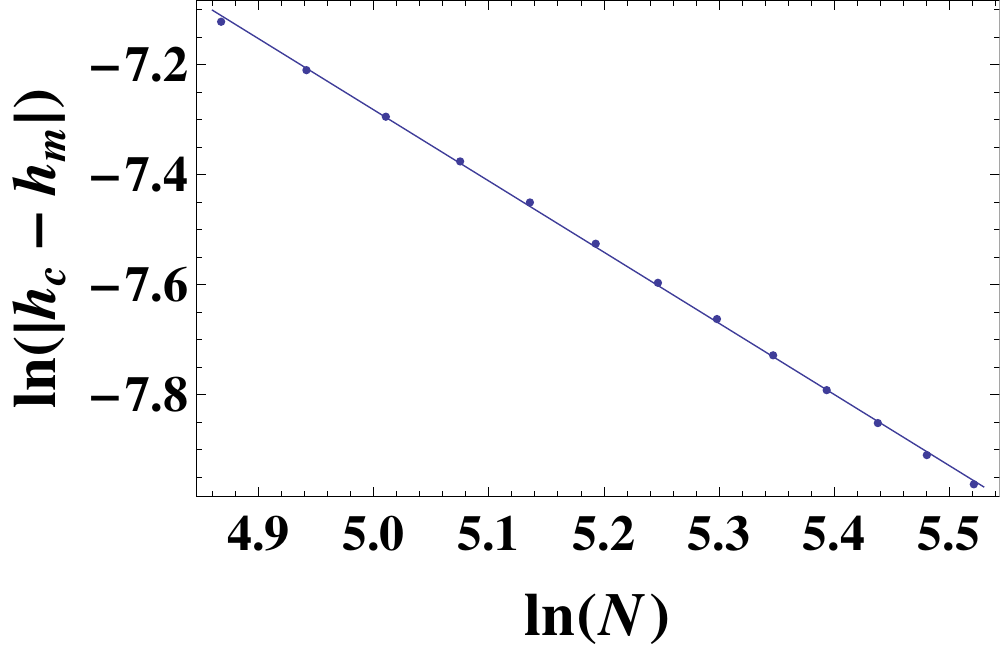}}
\def\lowerinsetent{\includegraphics[height=2cm]{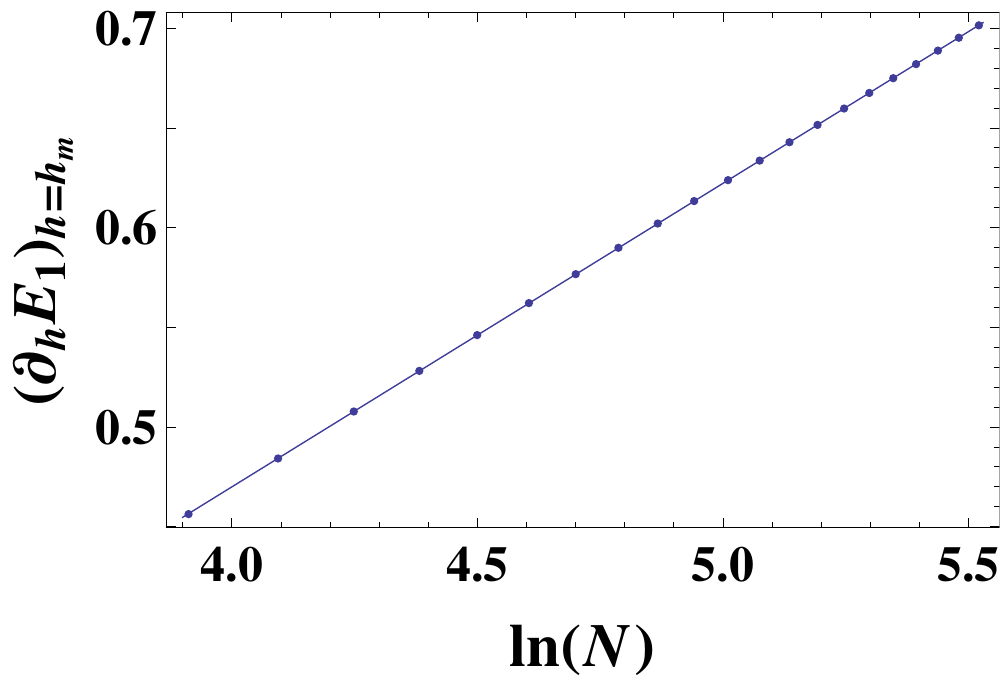}}
\stackinset{r}{8pt}{t}{7pt}{\upperinsetent}{%
\stackinset{r}{8pt}{b}{35pt}{\lowerinsetent}{%
\includegraphics[height=5.5cm]{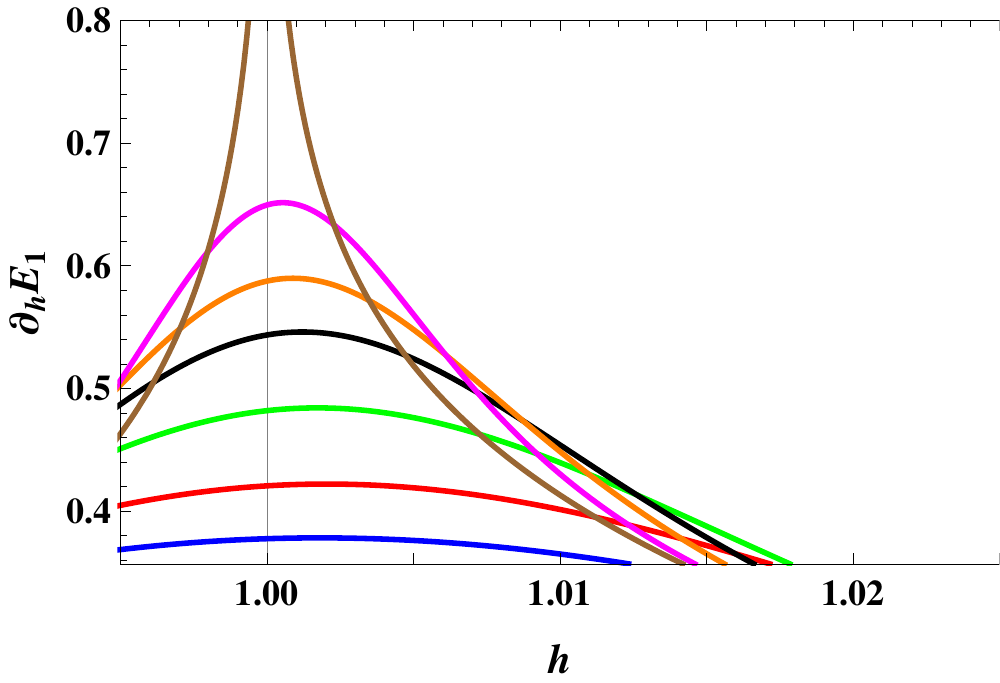}%
}}
\caption{First derivative of the nearest-neighbor discord of response (upper panel) and of the nearest-neighbor entanglement of response (lower panel) in symmetry-preserving ground states, as functions of the external field $h$ in proximity of the critical point, in the case of $\gamma=0.5$, for different chain lengths $N$. Blue curve: $N=30$; red curve: $N=40$; green curve: $N=60$; black curve: $N=90$; orange curve: $N=120$; magenta curve: $N=180$; brown curve: $N=\infty$. In both panels, the solid vertical line represents the critical point. In both panels, the upper inset shows the dependence on the chain size $N$ of the renormalized critical point $h_m$. The lower inset in the upper panel displays the dependence on the chain size $N$ of the value attained at the renormalized critical point $h_m$ by the first derivative of the nearest-neighbor discord of response. The lower inset in the lower panel displays the same dependence for the nearest-neighbor entanglement of response.}
\label{fig:firstdersymentandquantversushnearcriticwithtwoinsets}
\end{figure}

When increasing the inter-spin distance $r$, the pairwise entanglement of response $E_r$ and discord of response $Q_r$ behave even more differently (see Fig.~\ref{fig:symnquantandentversushatvariousr}). $E_r$ dramatically drops to zero as $r$ increases, except in a small region
around the factorizing field $h=h_f$ that gets smaller and smaller as $r$ increases, in agreement with the findings of Ref.~\cite{ABFPTV2006}. On the other hand, while in the disordered and critical phases $Q_r$ vanishes as $r$ increases, in the ordered phase $Q_r$ survives even in the limit of
infinite $r$. Indeed, in both the disordered and critical phases, and when $r$ goes to infinity, the only nonvanishing one-body and two-body correlation functions in the symmetry-preserving ground states are
$\langle \sigma_i^z \rangle$ and $\langle \sigma_i^z \sigma_{i+r}^z\rangle$, so that the two-body reduced state can be written as
a classical mixture of eigenvectors of $\sigma_i^z \sigma_{i+r}^z$. On the other hand, in the ordered phase, also the two-body correlation function $\langle \sigma_i^x \sigma_{i+r}^x\rangle$ appears, while $\langle \sigma_i^x \rangle$ vanishes due to symmetry preservation, thus preventing the two-body marginal of the symmetry-preserving ground state from being a mixture of classical states.

The long range nature of the pairwise discord of response not only tells us that quantum correlations beyond entanglement cannot be monogamous, at variance with entanglement itself~\cite{CKW2000,OV2006}, but also reveals the unavoidable quantum nature of the symmetry-preserving ground states in the ordered phase. Indeed, symmetry-preserving ground states are, in general, entangled coherent superpositions of the symmetry-breaking ordered ground states, and the latter are the most classical ones among all possible ground states, as recently proven in Refs.~\cite{CFGI2014,HGI2015}. At the factorizing fields, the symmetry-preserving ground states are maximally entangled coherent superpositions (Schroedinger cats).

Let us now analyze the behavior of the nearest-neighbor entanglement of response $E_1$ and discord of response $Q_1$ in close proximity to the quantum critical point $ h_c=1$.
Fig.~\ref{fig:firstdersymentandquantversushnearcriticwithtwoinsets} shows that both $\partial_hE_1$ and $\partial_hQ_1$ manifest a logarithmic divergence at the critical point, for any non zero anisotropy $\gamma$. However, while in the entanglement case such logarithmic divergence is always positive, in the discord case it can be either positive or negative, depending on the anisotropy $\gamma$.


\begin{figure}[t]
\def\insetquant{\includegraphics[height=2.0cm]{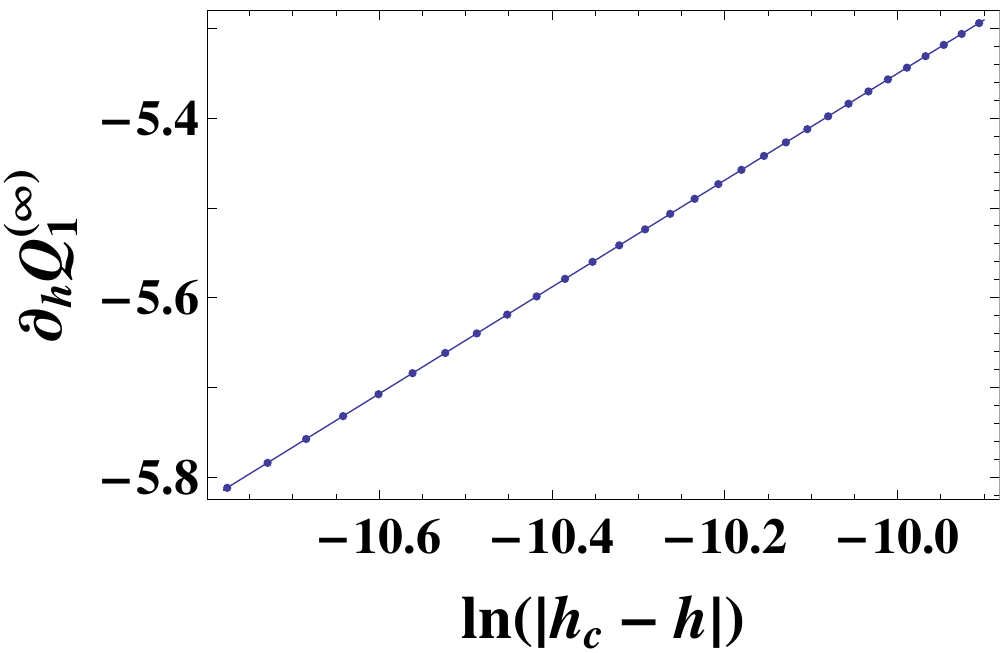}}
\stackinset{c}{}{d}{35pt}{\insetquant}{\includegraphics[height=5.3cm]{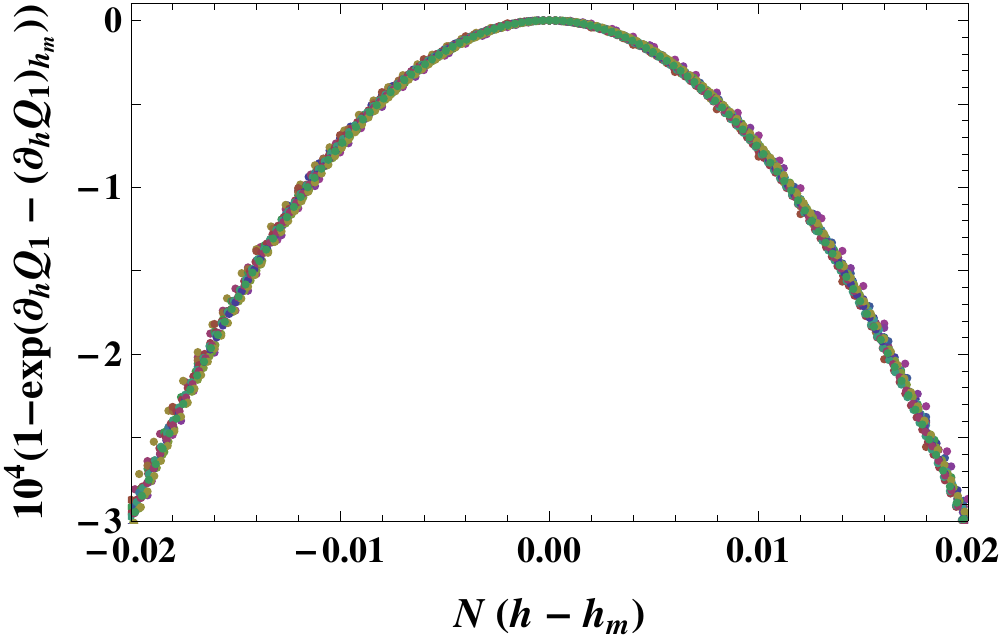}}
\def\insetent{\includegraphics[height=2.0cm]{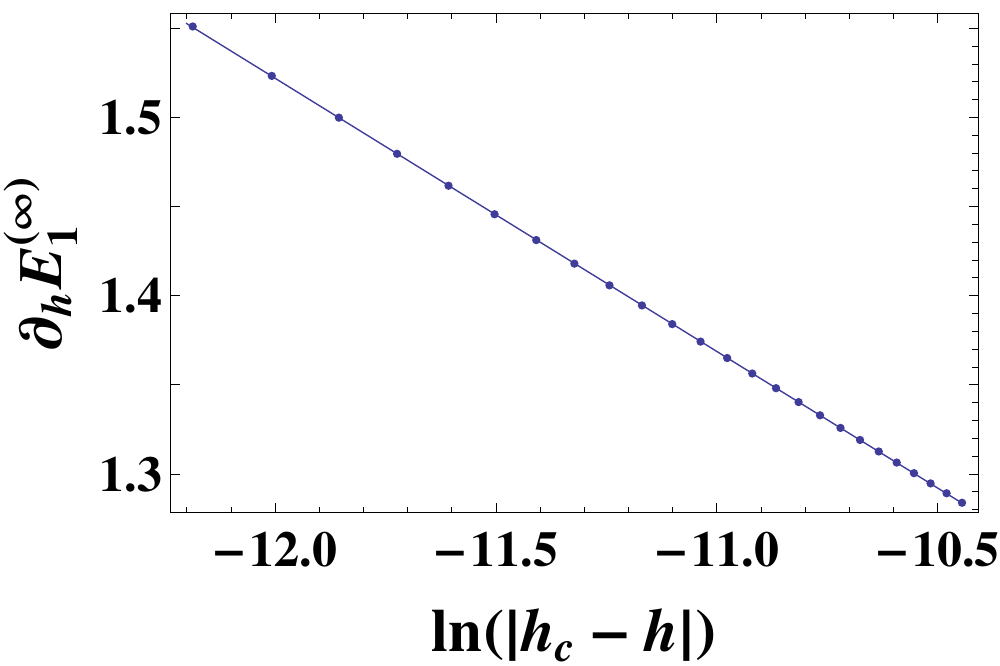}}
\stackinset{c}{}{t}{15pt}{\insetent}{\includegraphics[height=5.5cm]{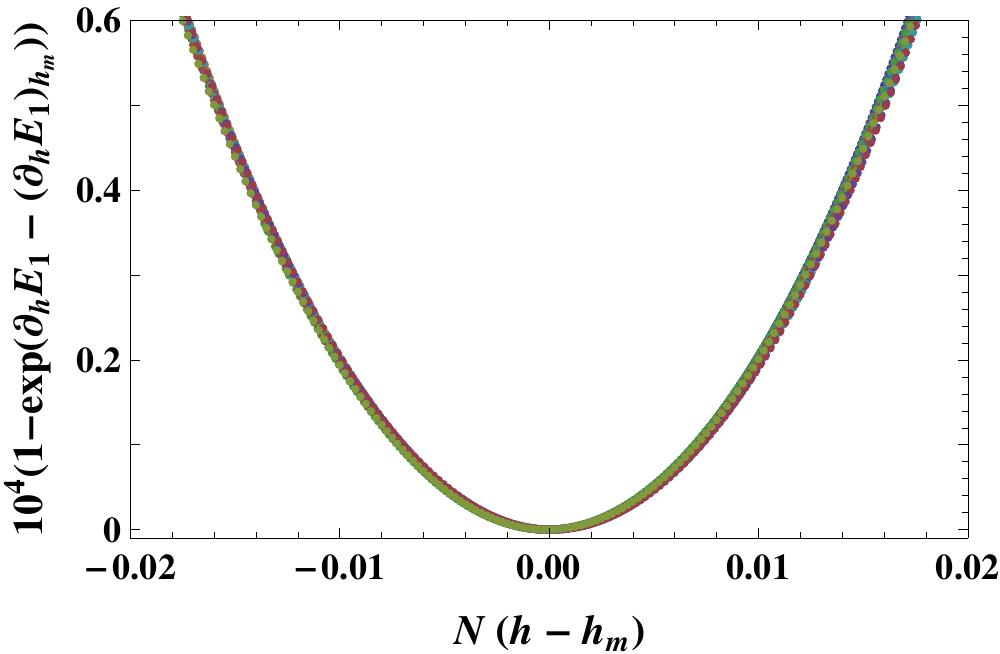}}
\caption{Finite size scaling of the first derivative of the nearest-neighbor discord of response (upper panel) and of the nearest-neighbor entanglement of response (lower panel) in symmetry-preserving ground states, for $\gamma=0.5$. The inset in the upper panel shows the first derivative of the nearest-neighbor discord of response in the thermodynamic limit as a function of the proximity to the critical point. The same in the inset in the lower panel, but for the nearest-neighbor entanglement of response.}
\label{fig:finitesizescalingsymquantandentnearcritic}
\end{figure}

By performing finite size scaling analysis, one can show that both $\partial_h E_1$ and $\partial_h Q_1$ allow for an accurate description
of the quantum phase transition occurring at $h_c=1$, providing us with the corresponding critical exponent $\nu$~\cite{B1983,S2000,OAFF2002,GH2013}.
In the following, for the sake of illustration, we obtain the critical exponent $\nu$ for the $XY$-model with intermediate anisotropy $\gamma=0.5$, although the same result applies to any non zero anisotropy. To do this, we need to suitably compare the two following behaviors~\cite{B1983}.

On the one hand, we have the dependence on the chain size $N$ of the value attained at the renormalized critical point $h_m$ by $\partial_h E_1$  (resp., $\partial_h Q_1$), where $h_m$ is its maximum (resp., minimum) in the close proximity of the critical point $h_c=1$ (see Fig.~\ref{fig:firstdersymentandquantversushnearcriticwithtwoinsets}):
\begin{eqnarray}\label{eq:maxfirstderversuslogN}
\left. \partial_h E_1^{(N)}\right|_{h_m} & = & 0.15 \ln N + const \; , \nonumber \\
\left. \partial_h Q_1^{(N)}\right|_{h_m} & = & - 0.59 \ln N + const \; .
\end{eqnarray}


On the other hand, we have the dependence on the proximity to the critical point $h_c=1$ of $\partial_h E_1$  (resp., $\partial_h Q_1$) in the thermodynamic limit (see insets of Fig.~\ref{fig:finitesizescalingsymquantandentnearcritic}):
\begin{eqnarray}\label{eq:firstderversusloghcminush}
\partial_h E_1^{(\infty)} & = & - 0.15 \ln\left|h_c-h\right| + const\; , \nonumber \\
\partial_h Q_1^{(\infty)} & = & 0.59 \ln\left|h_c-h\right| + const\; .
\end{eqnarray}

According to the scaling ansatz relative to the case of a logarithmic divergence~\cite{B1983}, the critical exponent $\nu$ is simply given by the opposite of the ratio between the pre-factors of the logarithms in Eq.~$(\ref{eq:firstderversusloghcminush})$ and Eq.~$(\ref{eq:maxfirstderversuslogN})$,
respectively. We thus obtain $\nu=1$, in agreement with the known fact that all the anisotropic $XY$ models belong to the Ising universality class. Moreover, Fig.~\ref{fig:finitesizescalingsymquantandentnearcritic} highlights the precision of the above finite size scaling analysis by showing that, via a proper scaling of the functions $\partial_h E_1$ and $\partial_h Q_1$~\cite{B1983}, it is possible to make all the data for different sizes $N$ collapse onto a single curve.

\begin{figure}[t]
\includegraphics[width=8cm]{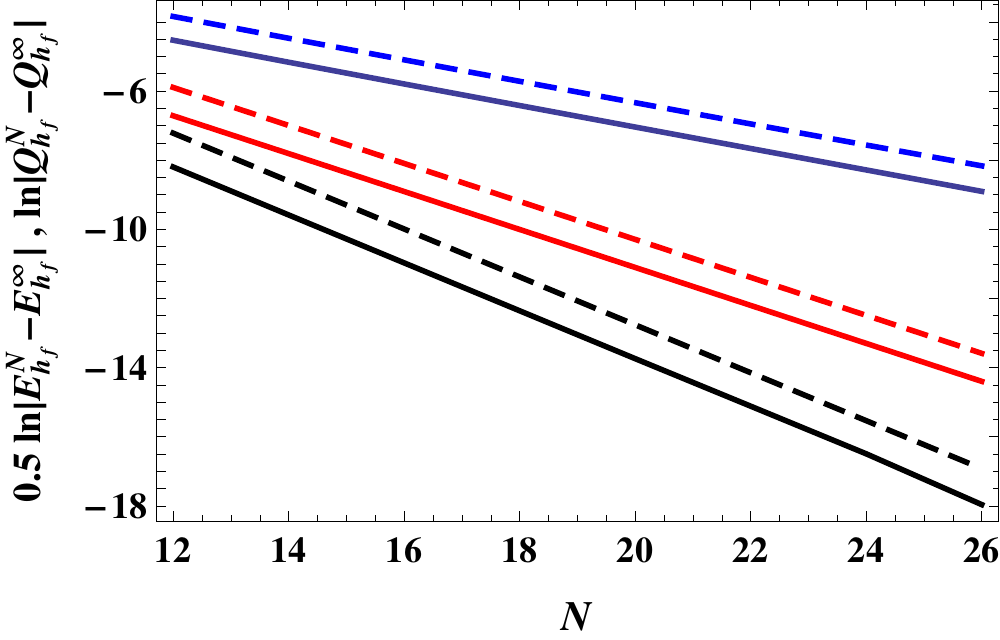}
\caption{Dependence on the chain size $N$ of the value attained at the factorizing field by the nearest-neighbor discord of response
(solid line) and nearest-neighbor entanglement of response (dashed line) for symmetry-preserving ground states, at different values of $\gamma$.
Blue lines: $\gamma=0.3$; red lines: $\gamma=0.5$; black lines: $\gamma=0.6$.}
\label{fig:finitesizescalingsymquantandentfactorization}
\end{figure}

We conclude this section by studying how the nearest-neighbor entanglement of response $E_1$ and the nearest-neighbor discord of response $Q_1$ scale with the system size $N$ in close proximity to the factorizing field $h = h_f=\sqrt{1-\gamma^2}$. In spite of the significant difference between $E_1$ and $Q_1$ at the factorizing field in the thermodynamic limit, as shown in Fig.~\ref{fig:symnearestquantandent}, their finite size scalings in the proximity of $h_f$ are extremely similar, as shown in Fig.~\ref{fig:finitesizescalingsymquantandentfactorization}. Indeed, both the entanglement and the discord scale with $N$ according to an exponential decay that is independent of the inter-spin distance $r$ and gets faster and faster as the anisotropy $\gamma$ increases. Interestingly, for any fixed value of the anisotropy $\gamma$, the decay rate of the entanglement of response is twice that of the corresponding rate for the discord of response.

\section{Symmetry-breaking ground states}\label{sec:symmmetrybrokengroundstate}

\begin{figure}[t]
\includegraphics[width=8cm]{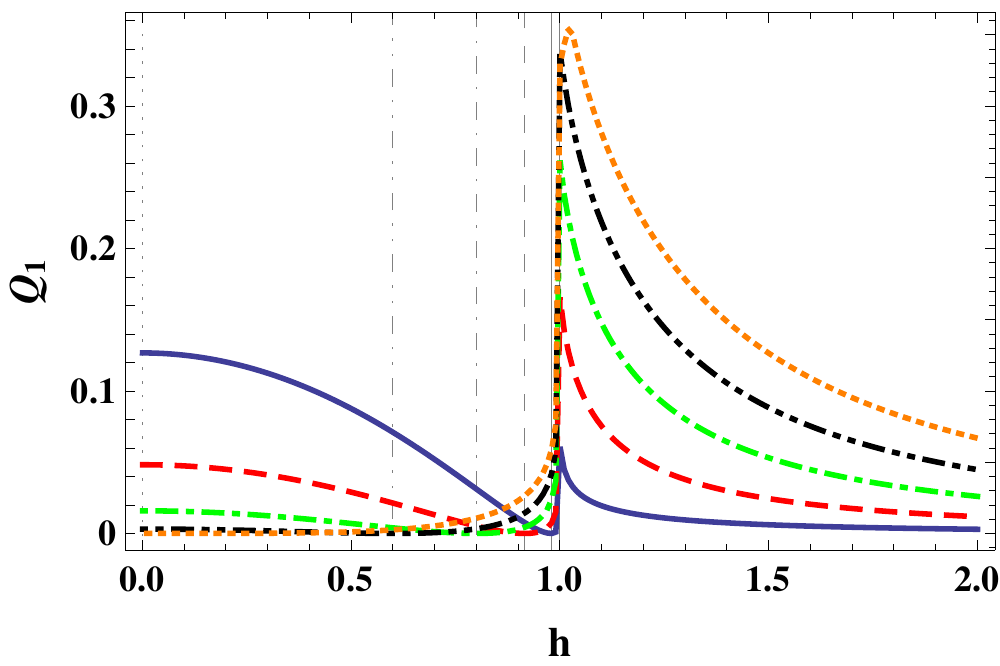}
\includegraphics[width=8cm]{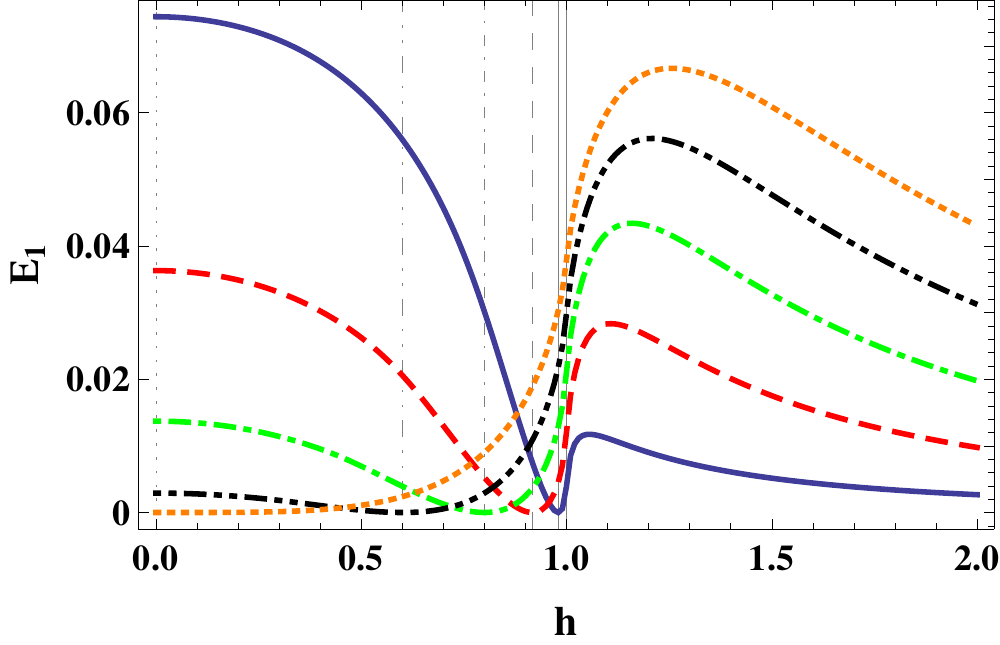}
\caption{Nearest-neighbor discord of response (upper panel) and nearest-neighbor entanglement of response (lower panel) in symmetry-breaking ground states as  functions of the external field $h$, for different values of the anisotropy $\gamma$. Solid blue curve: $\gamma=0.2$; dashed red curve: $\gamma=0.4$;
dot-dashed green curve: $\gamma=0.6$; double-dot-dashed black curve: $\gamma=0.8$; dotted orange curve: $\gamma=1$. In both panels, to each of these curves, there corresponds a vertical line denoting the associated factorizing field $h_f$. The rightmost vertical line denotes the critical point.}
\label{fig:symbrokennearestquantandent}
\end{figure}
\begin{figure}[hbtp]
\includegraphics[width=8cm]{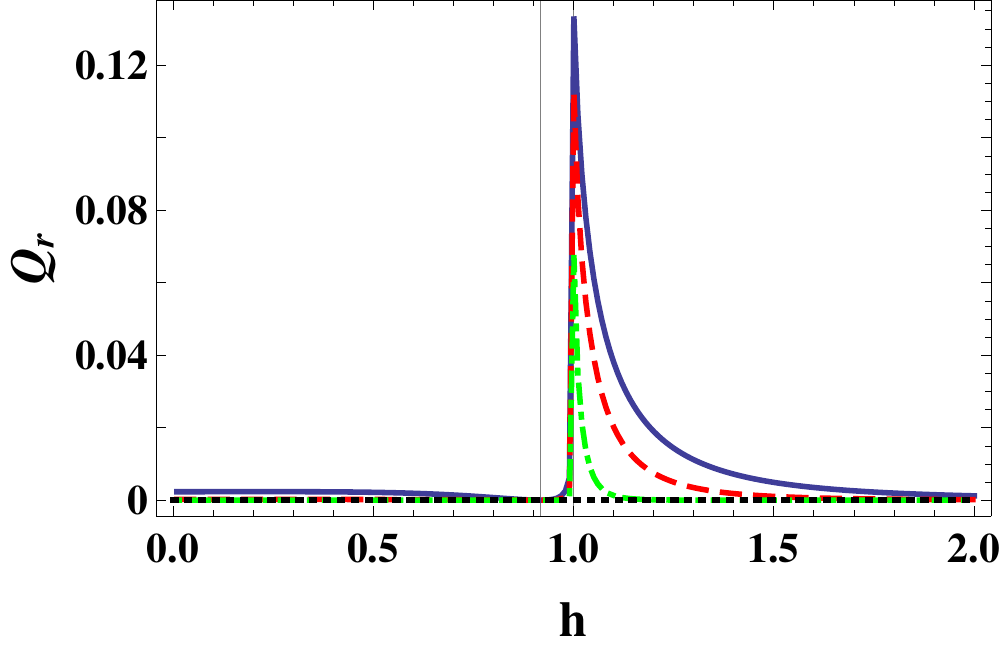}
\includegraphics[width=8cm]{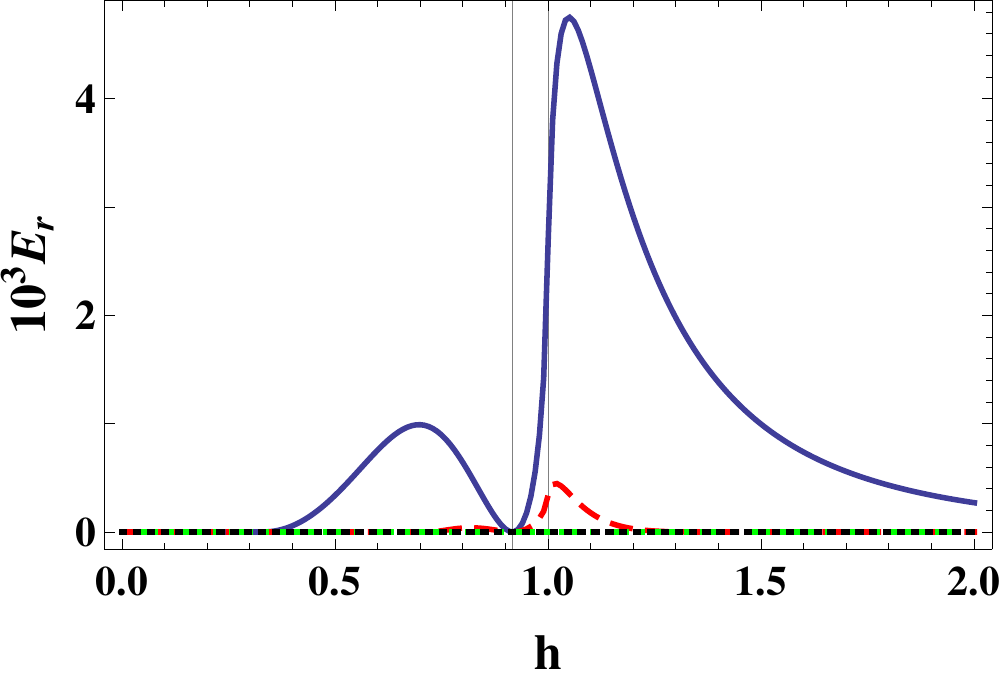}
\caption{Two-body discord of response (upper panel) and two-body entanglement of response (lower panel) in symmetry-breaking ground states as functions of the external field $h$, at $\gamma=0.4$, for different inter-spin distances $r$. Solid blue curve: $r=2$; dashed red curve: $r=3$;
dot-dashed green curve: $r=8$; dotted black curve: $r=\infty$. In both panels, the two solid vertical lines correspond, respectively, to the factorizing field (left) and to the critical field (right).}
\label{fig:symbrokennquantandentversushatvariousr}
\end{figure}

In this section we move the focus of the comparison between two-body entanglement of response and discord of response from symmetry-preserving to symmetry-breaking ground states. Spontaneous symmetry breaking manifests itself in the thermodynamic limit, in the ordered phase $h<h_c=1$ and for any non zero anisotropy $\gamma$, so that hereafter we will restrict the region of the phase space under investigation accordingly.


Fig.~\ref{fig:symbrokennearestquantandent} shows that, as soon as symmetry breaking is taken into account, the nearest-neighbor discord of response $Q_1$ becomes discontinuous at the critical point $h_c=1$, whereas the first derivative of the nearest-neighbor entanglement of response $\partial_h E_1$ still diverges logarithmically. In other words, only the discord of response is affected by symmetry breaking at the critical point $h_c=1$. In fact, according to Ref.~\cite{OPM2006}, the concurrence and, consequently, the two-body entanglement of response, attain the same value for any $h\geq h_f$ both in the symmetry-preserving and symmetry-breaking ground states. Otherwise, if $h<h_f$, there is a slight enhancement in the pairwise entanglement of response in the symmetry-breaking ground states compared to the corresponding symmetry-preserving ones. Conversely, in general, the pairwise discord of response undergoes a dramatic suppression in the entire ordered phase $h<h_c$ when moving from symmetry-preserving to symmetry-breaking ground states.

Overall, the quantum correlations between two neighboring spins decrease significantly in the entire ordered phase when symmetry breaking is taken into account, and are almost entirely made up by contributions due to entanglement. In particular, at the factorizing field $h_f$, both the entanglement of response and the discord of response vanish. Indeed, we recall that the factorizing field $h_f$ owes its name to the two symmetry-breaking ground states that are completely separable (product) at such value of the external magnetic field.

\begin{figure}[t]
$
\begin{array}{cc}
\includegraphics[height=2.75cm]{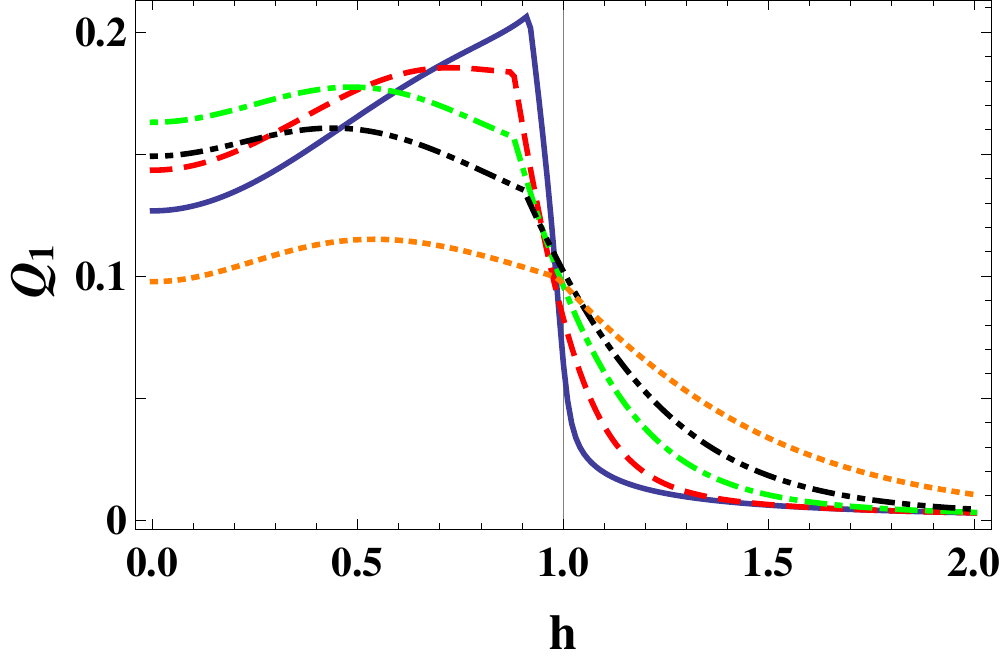}
\includegraphics[height=2.75cm]{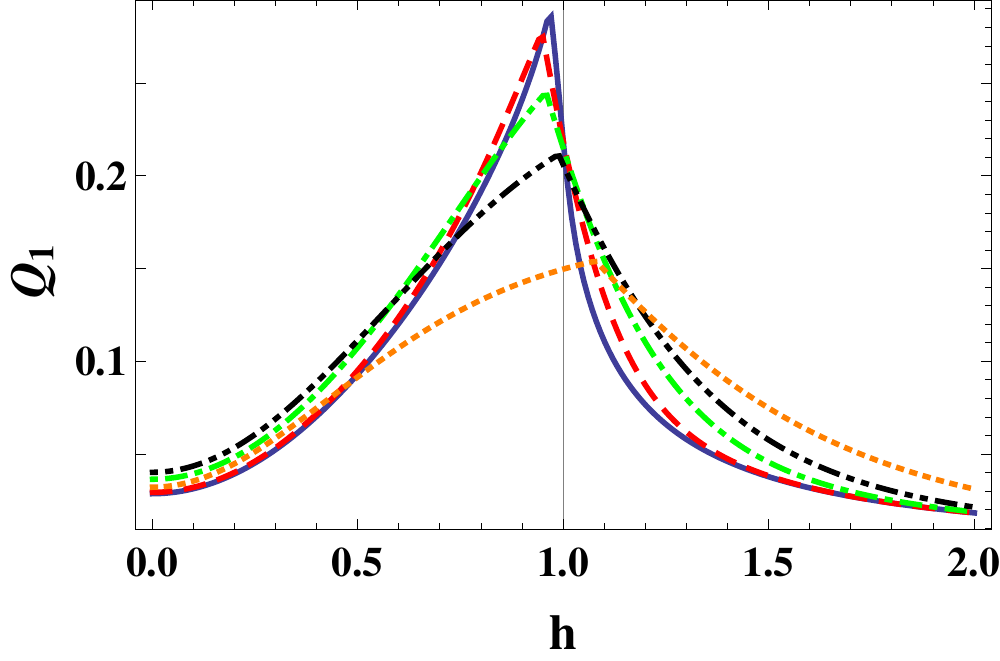}
\end{array}
$
$
\begin{array}{cc}
\includegraphics[height=2.75cm]{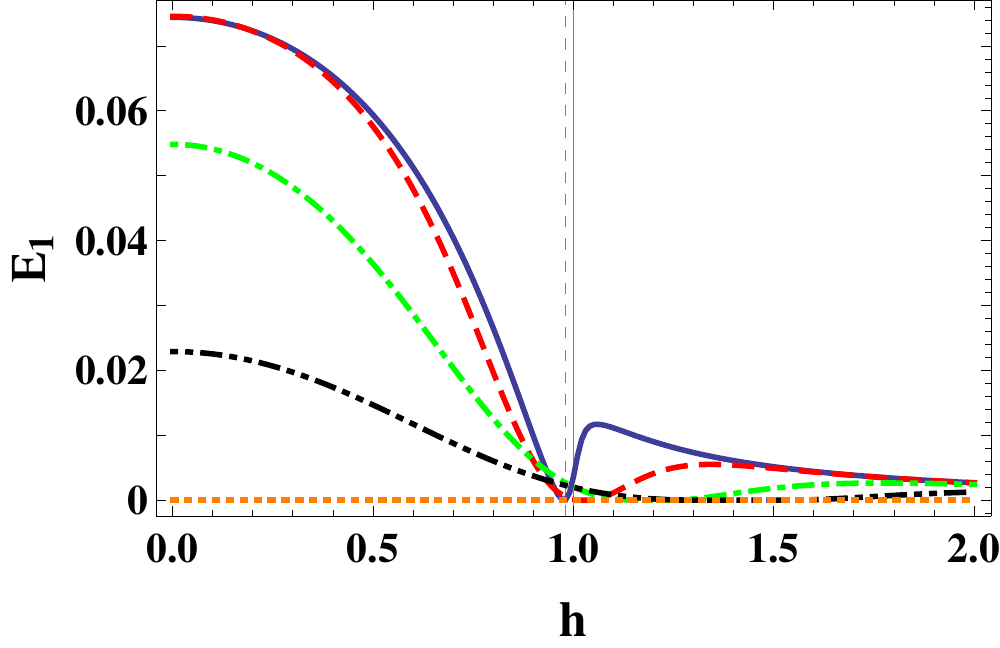}
\includegraphics[height=2.75cm]{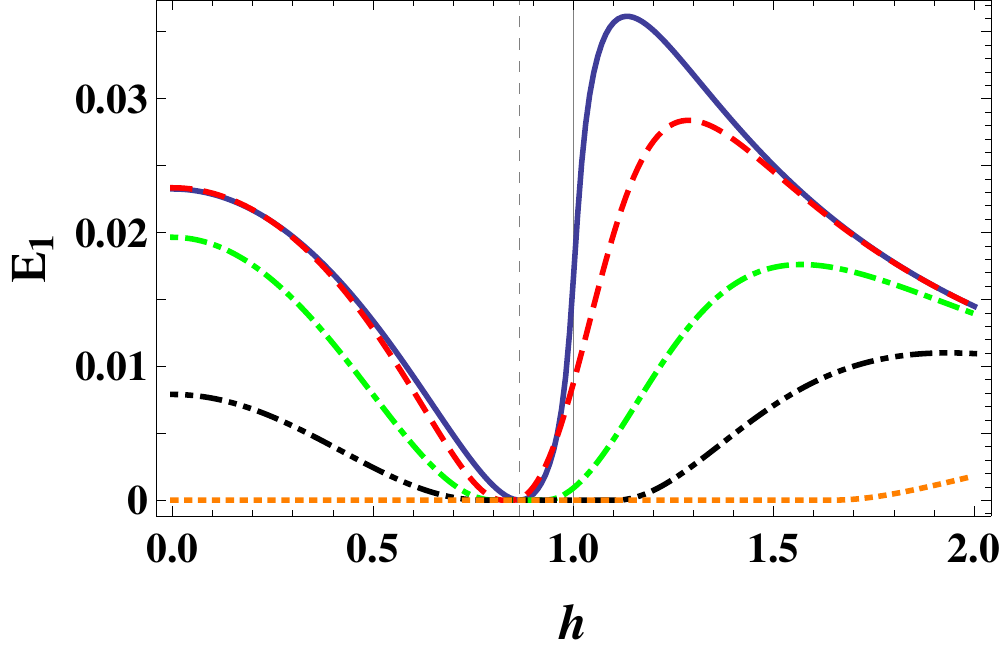}
\end{array}
$
\caption{Nearest-neighbor discord of response (leftmost upper panel) and nearest-neighbor entanglement of response (leftmost lower panel) for thermal states, as functions of the external field $h$, in the thermodynamic limit, with $\gamma=0.2$, for different values of the temperature $T$. Solid blue curve: $T=0.01$; dashed red curve: $T=0.1$; dot-dashed green curve: $T=0.2$; double-dot-dashed black curve: $T=0.3$; dotted orange curve: $T=0.5$. Rightmost panels: same, but with $\gamma=0.5$. In all panels, the vertical solid line corresponds to the critical point. In the lower panels, the dashed vertical lines correspond to the factorizing field.}
\label{fig:thermalentandquantversushatvariousT}
\end{figure}

\begin{figure}[t]
\includegraphics[width=7.5cm]{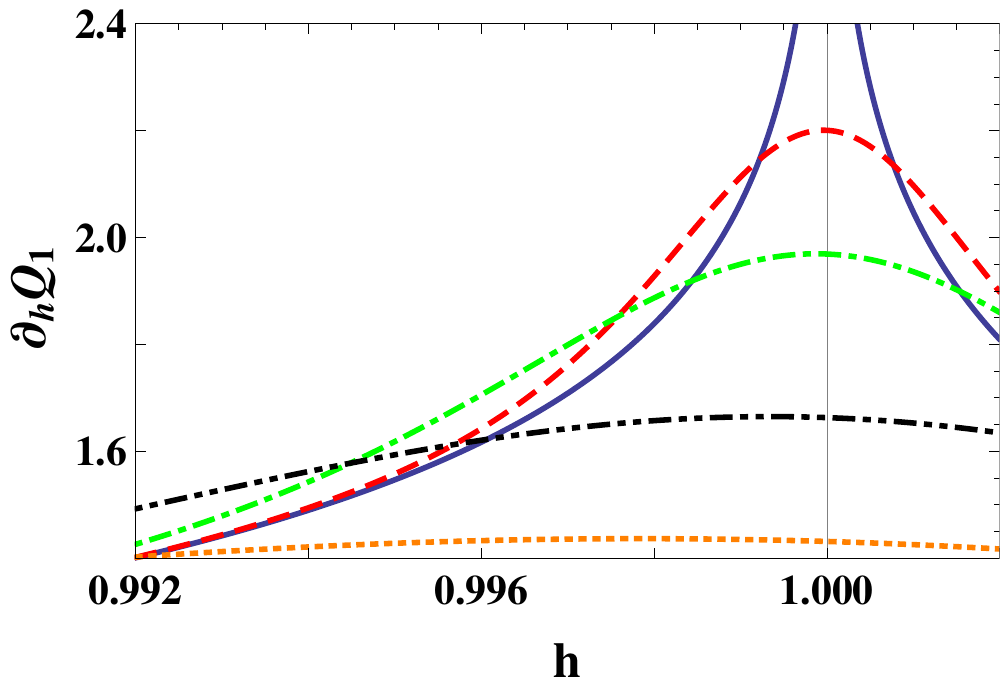}
\includegraphics[width=8cm]{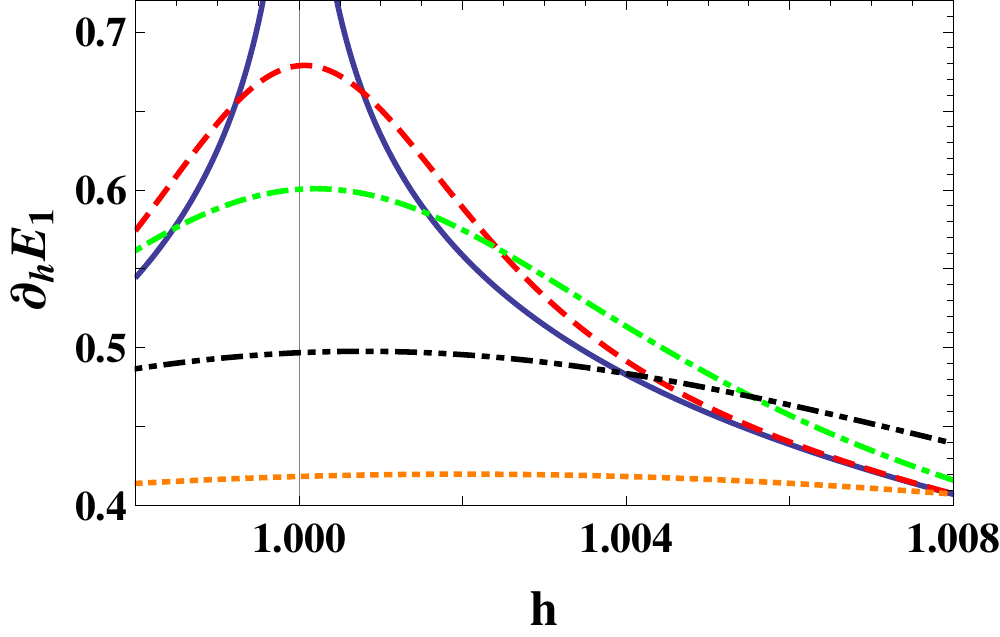}
\caption{First derivative of the nearest-neighbor discord of response (upper panel) and of the nearest-neighbor entanglement of response (lower panel) for thermal states, in the thermodynamic limit, as functions of the external field $h$, with $\gamma=0.9$, for different values of the temperature $T$. Solid blue curve: $T=0$; dashed red curve: $T=0.001$; dot-dashed green curve: $T=0.002$; double-dot-dashed black curve: $T=0.005$; dotted orange curve:$T=0.01$. In both panels, the solid vertical line represents the critical point.} \label{fig:firstderthermalentandquantversushatvariousr}
\end{figure}

\begin{figure}[t]
\includegraphics[width=8cm]{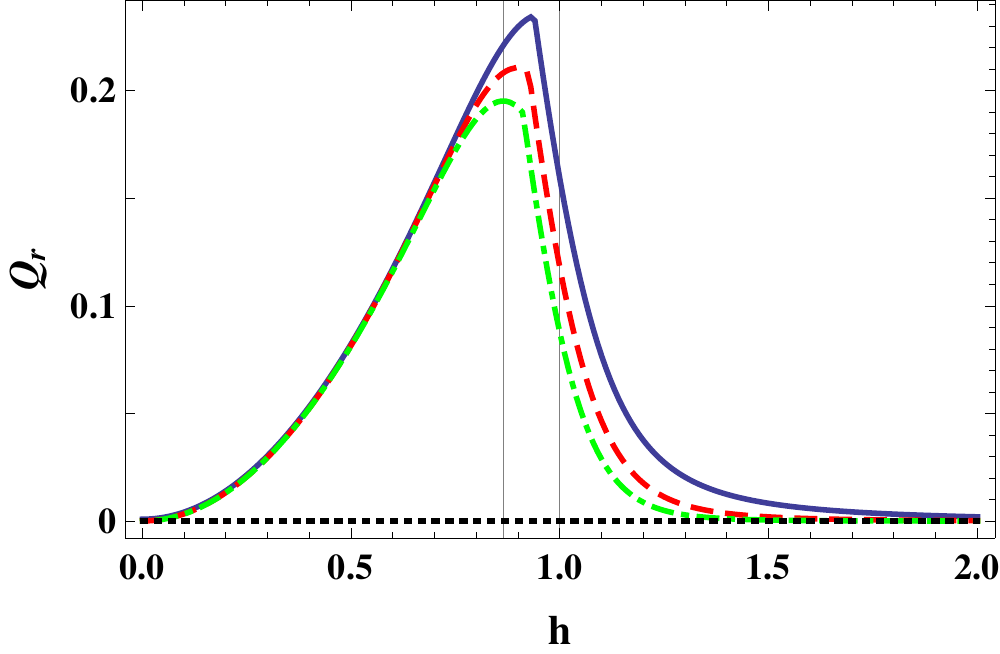}
\includegraphics[width=8cm]{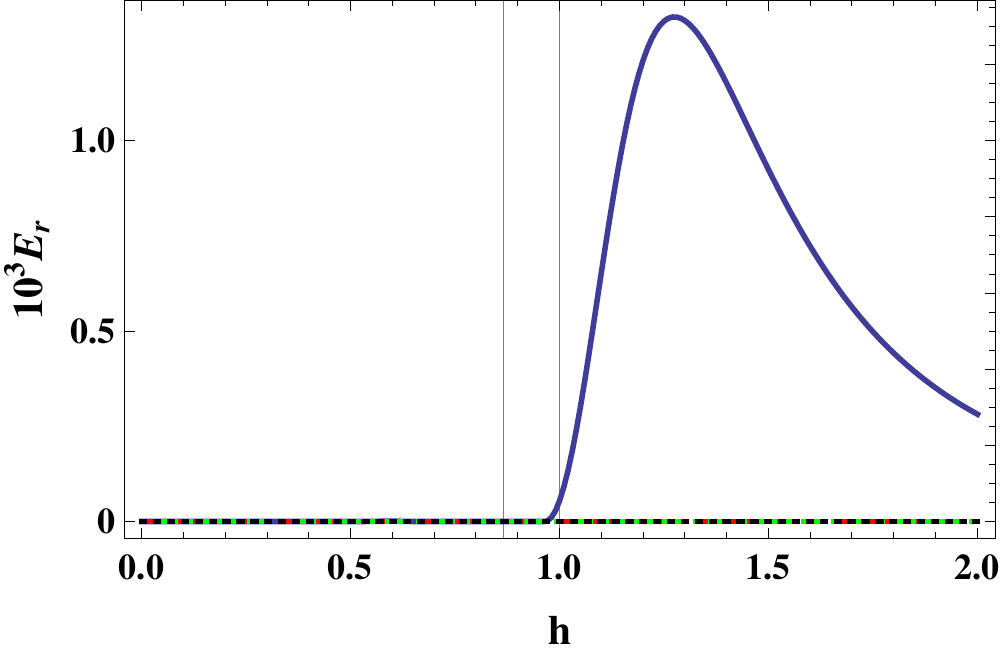}
\caption{Two-body discord of response (upper panel) and two-body entanglement of response (lower panel) for thermal states at temperature $T=0.1$, in the thermodynamic limit, as functions of the external field $h$, with $\gamma=0.5$, for different values of the inter-spin distance $r$. Solid blue curve: $r=2$; dashed red curve: $r=3$; dot-dashed green curve: $r=4$; dotted black curve: $r=\infty$. In both panels, the two solid vertical lines correspond, respectively, to the factorizing field (left) and to the critical field (right).}
\label{fig:thermalentandquantversushatvariousr}
\end{figure}

Considering the dependence on the inter-spin distance, we observe that the pairwise discord of response loses its long-range nature when moving from symmetry-preserving to symmetry-breaking ground states (see Fig.~\ref{fig:symbrokennquantandentversushatvariousr}). More precisely, both the pairwise entanglement of response and the pairwise discord of response vanish asymptotically with increasing inter-spin distance.

In the case of the pairwise entanglement of response, this result is again due to the monogamy of the squared concurrence~\cite{CKW2000,OV2006}. In the case of the pairwise discord of response, it is instead due to the fact that not only the correlation function $\langle \sigma_i^x \sigma_{i+r}^x\rangle$ but also $\langle \sigma_i^x\rangle$ and $\langle \sigma_i^x \sigma_{i+r}^z\rangle$ are nonvanishing in the limit of infinite inter-spin distance $r$. This feature allows to write any two-spin reduced density matrix obtained from the symmetry-breaking ground states as a classical mixture of eigenvectors of $O_i O_{i+r}$,
where $O_i$ is an Hermitian operator defined on the $i$-th site as $O_i= \cos \beta \sigma_i^z + \sin \beta \sigma_i^x$ with $\tan \beta= \frac{\langle \sigma_i^x\rangle}{\langle \sigma_i^z\rangle}$.

Notwithstanding its dramatic reduction in the symmetry-breaking sector, the pairwise discord of response remains always larger than or equal to the corresponding pairwise entanglement of response for any value of $h$, $\gamma$ and $r$, as expected for a proper and reliable measure of quantum correlations more general than entanglement. We note in passing that this feature
is lost when instead of considering the trace distance, which is contractive under CPTP dynamical maps, we adopt the non-contractive Hilbert-Schmidt distance. Indeed, the Hilbert-Schmidt based discord of response becomes smaller than the corresponding entanglement of response for some values of the external field $h<h_f$. This stands as a fundamental physical illustration of the fact that the Hilbert-Schmidt distance cannot allow for a {\em bona fide} quantification of quantum correlations.

\section{Thermal states}\label{sec:thermalstate}

So far we have focused our analysis on the ground states of the $XY$ models, be them symmetry-preserving or symmetry-breaking. We will now consider $XY$ models in thermal equilibrium with a bath at finite temperature.

\begin{figure}[t]
$
\begin{array}{cc}
\includegraphics[height=2.75cm]{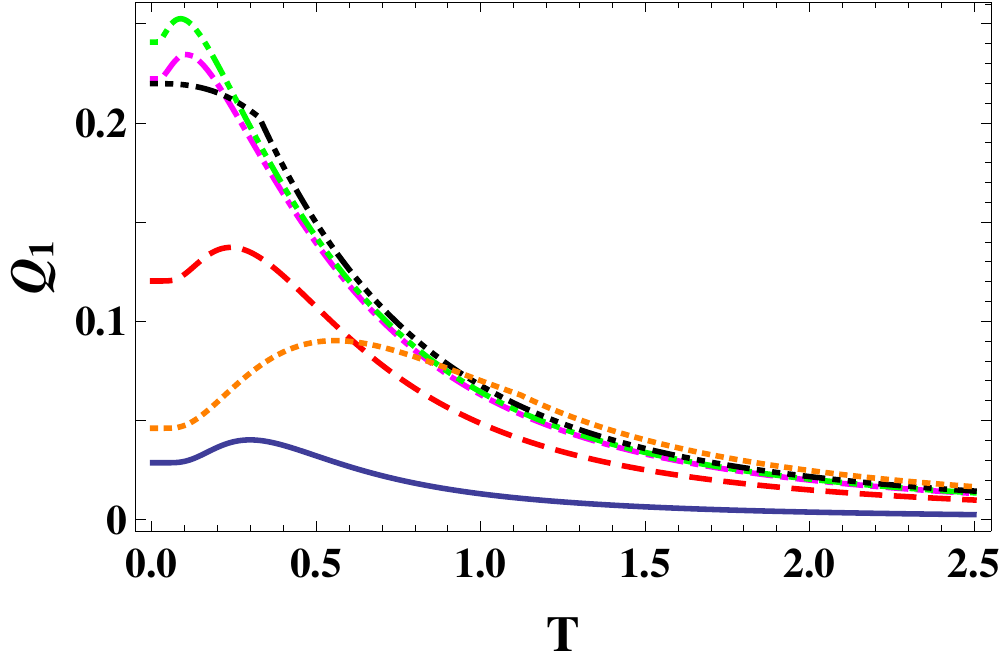}
\includegraphics[height=2.75cm]{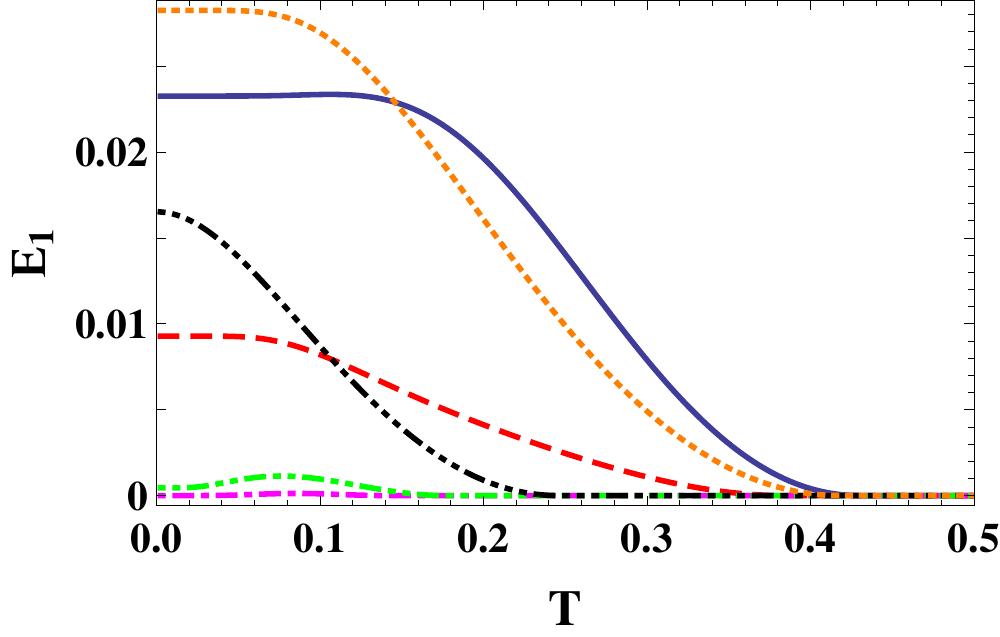}
\end{array}
$
$
\begin{array}{cc}
\includegraphics[height=2.75cm]{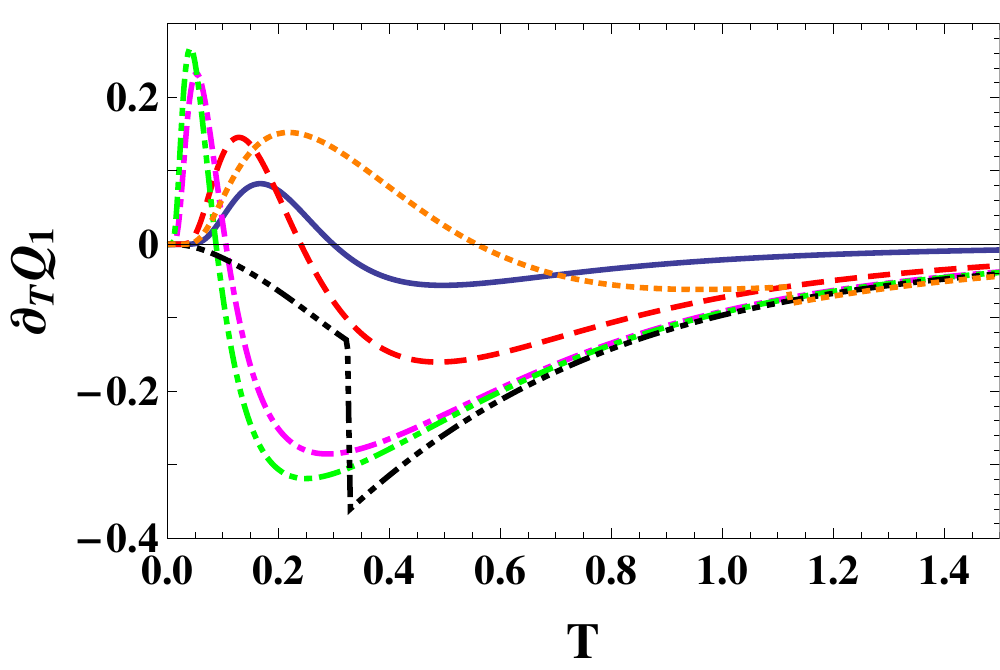}
\includegraphics[height=2.75cm]{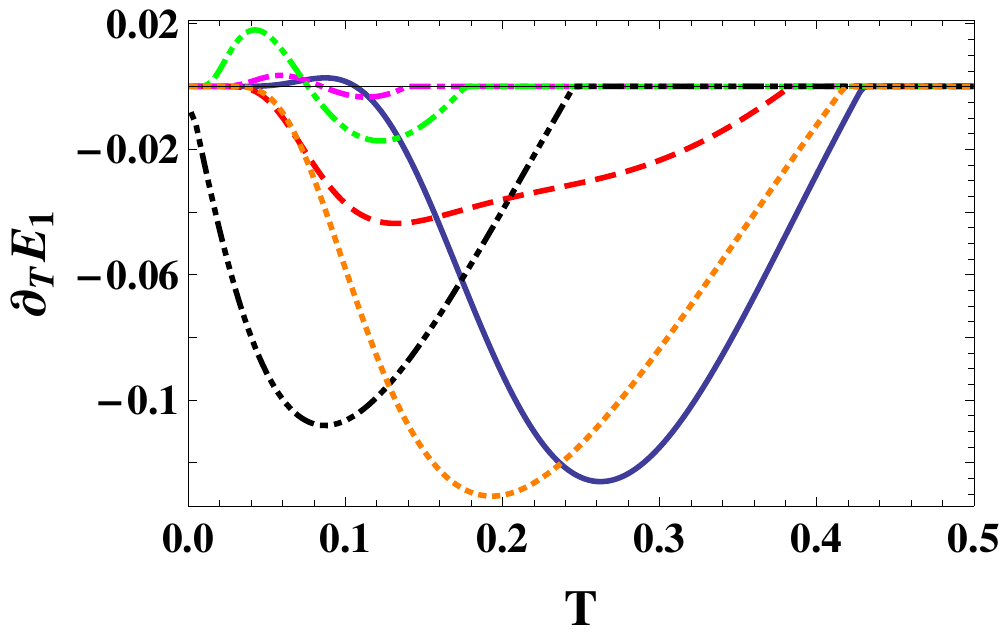}
\end{array}
$
\caption{Nearest-neighbor discord of response (leftmost upper panel) and its first derivative (leftmost lower panel) for thermal states, in the
thermodynamic limit, as functions of the temperature $T$, with $\gamma=0.5$, for different values of the external field $h$. Solid blue line: $h=0$; dashed red line: $h=0.6$; dot-dashed magenta line: $h=h_f$; double-dot-dashed green line: $h=0.9$; triple-dot-dashed black line: $h=h_c=1$; dotted orange line: $h=1.4$. The same in the rightmost panels, but for the nearest-neighbor entanglement of response.
}\label{fig:thermalentandquantversusTatvarioush}
\end{figure}

The behavior of the nearest-neighbor entanglement of response $E_1$ and of the nearest-neighbor discord of response $Q_1$, as well as that of their first derivatives, as functions of the external field, are plotted in
Figs.~\ref{fig:thermalentandquantversushatvariousT} and~\ref{fig:firstderthermalentandquantversushatvariousr} for different values of the temperature $T$.
The appearance of thermal effects has a rounding off effect that removes all singularities in correspondence of
the critical point $h_c$. Indeed, a sharp quantum phase transition can occur only at zero temperature. Specifically, as soon as the
temperature $T$ increases from zero to some finite value, the zero temperature singularity of $\partial_h E_1$ at the critical point $h_c$ is smoothed into
a maximum of $\partial_h E_1$ localized at a value of the external field $h$ higher than $h_c$. Moreover, the more the temperature increases,
the more this maximum is shifted away from the critical point $h_c$ and the corresponding value of $\partial_h E_1$ decreases.
Similarly, the divergence of $\partial_h Q_1$ at the critical point $h_c$ is replaced by either a minimum or a maximum of
$\partial_h Q_1$ (depending on $\gamma$) at a value of the external field $h$ lower than $h_c$. Furthermore, the higher $T$, the more this
extremal point moves away from the critical point $h_c$ and the corresponding absolute value of $\partial_h Q_1$ decreases.

Obviously, thermal effects also remove and/or distort the ground-state factorization phenomenon, that occurs at zero temperature for $h_f=\sqrt{1-\gamma^2}$. Indeed, Fig.~\ref{fig:thermalentandquantversushatvariousT} shows that, as the temperature varies, $h_f$ either belongs to a region where $E_1$ vanishes identically, or is a regular point at which $E_1$ is strictly nonzero.
Accordingly, as soon as the temperature $T$ increases from zero to some finite value, the discords evaluated for different inter-spin distances do not coincide anymore when $h=h_f$ (see Fig.~\ref{fig:thermalentandquantversushatvariousr}).

\begin{figure}[!t]

$
\begin{array}{cc}
\includegraphics[height=2.75cm]{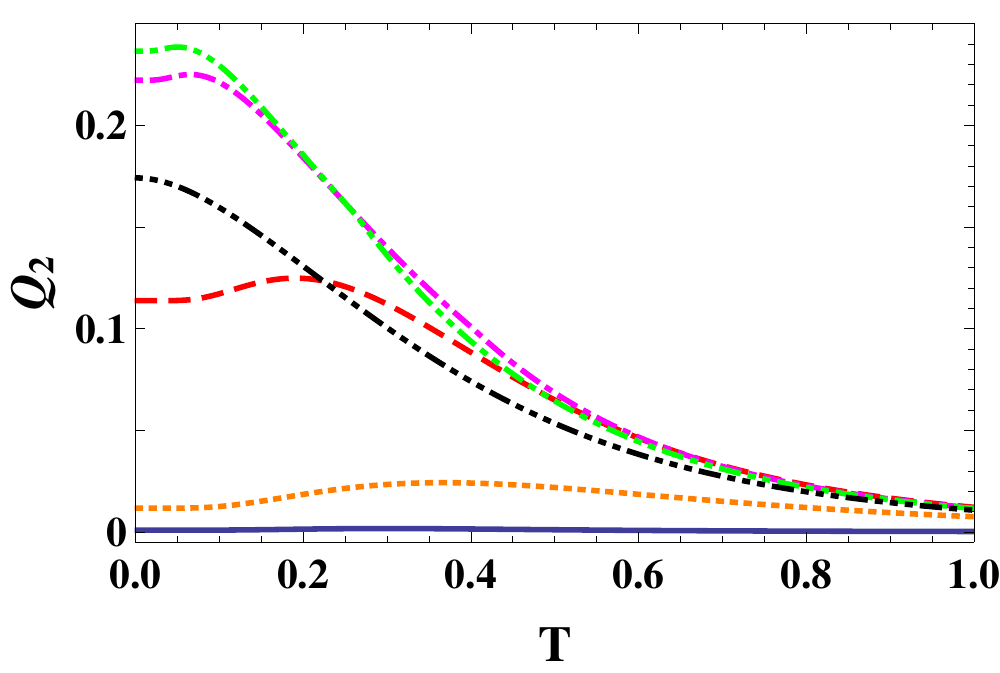}
\includegraphics[height=2.75cm]{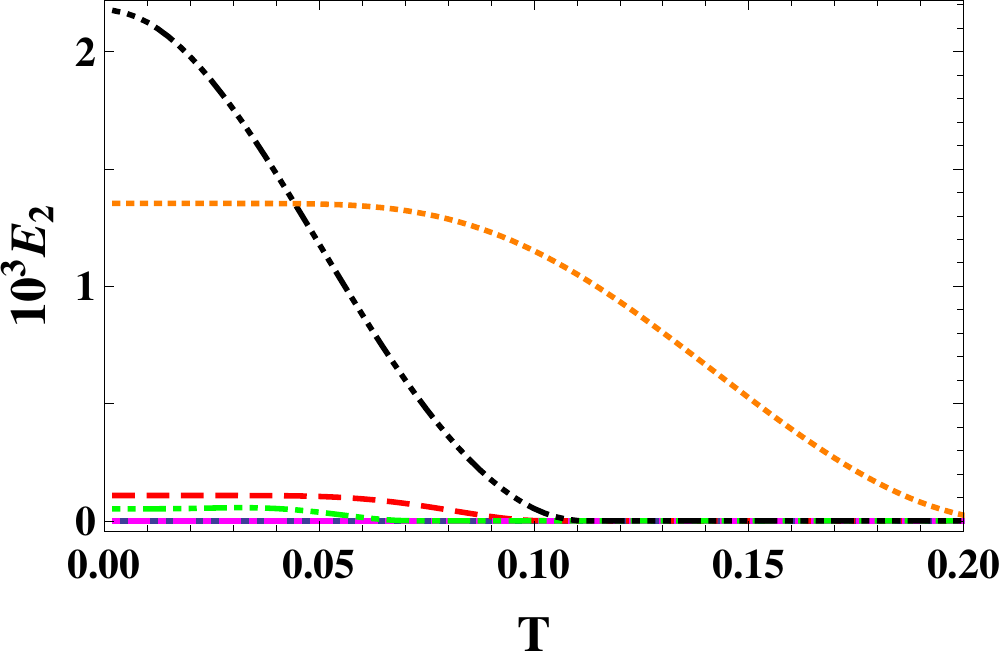}
\end{array}
$
$
\begin{array}{cc}
\includegraphics[height=2.75cm]{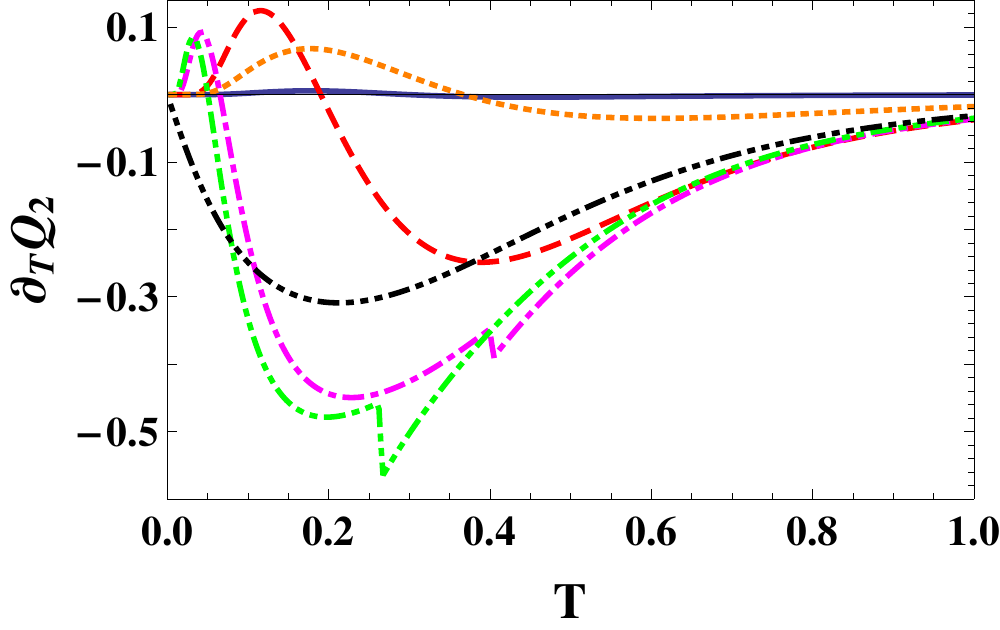}
\includegraphics[height=2.75cm]{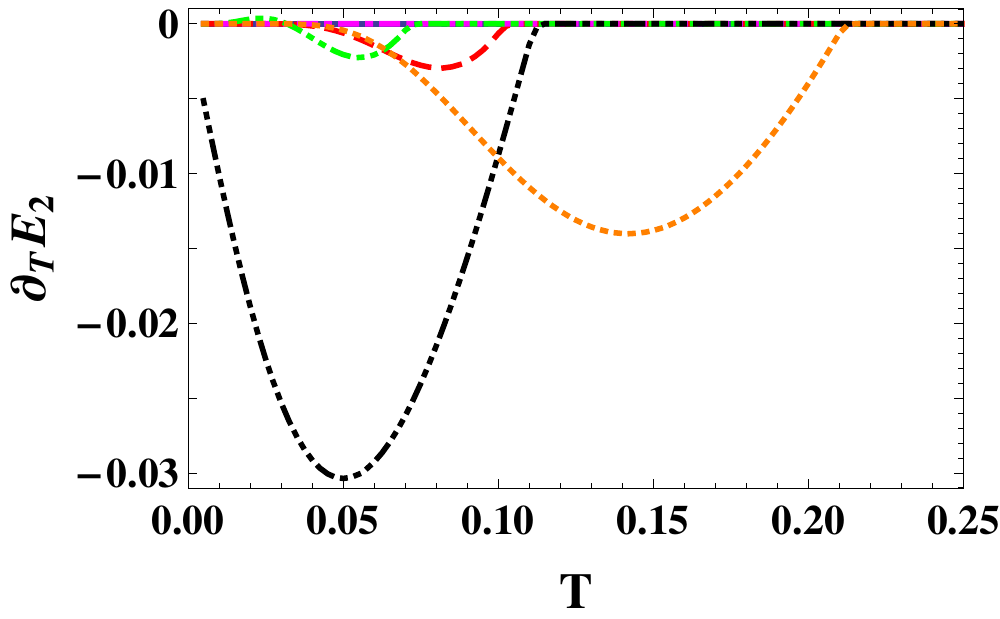}
\end{array}
$
\caption{Next-nearest-neighbor discord of response (leftmost upper panel) and its first derivative (leftmost lower panel) for thermal states, in the thermodynamic limit, as functions of the temperature $T$, in the case of $\gamma=0.5$, for different values of the external field $h$.  Solid blue line: $h=0$; dashed red line: $h=0.6$; dot-dashed magenta line: $h=h_f$; double-dot-dashed green line: $h=0.9$; triple-dot-dashed black line: $h=h_c=1$; dotted orange line: $h=1.4$. The same in the rightmost panels, but for the next-nearest-neighbor entanglement of response.
}\label{fig:thermalnextnearestentandquantversusTatvarioush}
\end{figure}

From Fig.~\ref{fig:thermalentandquantversushatvariousT} it also emerges that, as the temperature $T$ increases,
the peaks corresponding to $E_1$ and $Q_1$ tend to flatten to zero quite differently, with the pairwise discord being more robust than the pairwise entanglement against thermal effects.
Interestingly, Figs.~\ref{fig:thermalentandquantversusTatvarioush} and
~\ref{fig:thermalnextnearestentandquantversusTatvarioush} show that, for any anisotropy $\gamma$, there exist some values of
the external field $h$ such that either the pairwise entanglement or the pairwise discord increase with the temperature. This contrasts the common intuition for which thermal effects can only be detrimental to quantum features. Furthermore, this surprising behavior appears mostly in the case of the
pairwise discord. Indeed, the latter increases with the temperature for many values of the external field $h$ and any sufficiently small inter-spin distance $r$, whereas the pairwise entanglement displays such behavior only for particular values of $h$ and, essentially, in the case of pairs of nearest-neighboring spins.

More precisely, for sufficiently large anisotropy $\gamma$, $Q_1$ increases with the temperature for all values of the external field $h$ except for those belonging to a small interval around and including the critical point $h_c=1$. For sufficiently small $\gamma$, $Q_1$ increases with the temperature for all values of the external field $h$ except for those belonging to an interval that lies below the critical point $h_c=1$ and shifts towards lower values of $h$ as $\gamma$ decreases. Consequently, there is no correspondence between the growth of the pairwise discord with the temperature and the occurrence of a gap in the energy spectrum between the ground state and the first excited state.

\section{Conclusions and outlook}\label{sec:conclusions}

In this paper, by resorting to a unifying approach to the quantification of bipartite quantum correlations based on local unitary operations, we have performed the first, direct and comprehensive, comparison between the two-body entanglement and two-body quantum discord in infinite $XY$ quantum spin chains, both in symmetry-preserving and symmetry-breaking ground states as well as in thermal states at finite temperature.

For symmetry-preserving ground states, we have shown that the pairwise entanglement captures only a modest portion of the total pairwise quantum correlations. This fact is trivially obvious at the factorizing field and quite intuitive for long-range inter-spin distances: in both cases, the pairwise entanglement vanishes.

Conversely, for symmetry-breaking ground states, we have shown that the pairwise quantum correlations are strongly suppressed in the whole ordered phase $h<h_c$, while the pairwise entanglement is either unchanged or undergoes a slight enhancement, thus contributing the largest amount to the total pairwise quantum correlations. When adopting the Hilbert-Schmidt distance, we have also found that the two-body discord of response can be even smaller than the corresponding entanglement, thus providing a fundamental physical illustration of the fact that the Hilbert-Schmidt distance, being non contractive under CPTP maps, does not allow for a proper quantification of quantum correlations.

For thermal states at finite temperature, we have shown that the pairwise discord of response is in general more robust
than the pairwise entanglement against thermal effects. Moreover, we have also shown that a surprising resilience to thermal effects can occur both for the pairwise discord and the pairwise entanglement, whereby these quantum features can, in some regions of the Hamiltonian parameters, increase with the temperature, although this behavior appears most enhanced in the case of the pairwise discord.

The fact that pairwise quantum correlations and pairwise entanglement can increase with the temperature in some regimes of the Hamiltonian parameters, together with the complex behavior of the maximum pairwise entanglement and discord as functions of the anisotropy are two puzzling features whose physical origin is at present not fully understood and thus deserves further investigation.

It is finally worth remarking that in order to directly compare entanglement and quantum discord on equal footing, one might resort to other unifying approaches to the quantification of entanglement and quantum correlation besides the one based on the formalism of local unitary operations that we have used in the present work. It would then be interesting to see what conclusions can be drawn by comparing entanglement and quantum correlations, e.g. within the unifying geometric approach. In such approach, entanglement and quantum discord are quantified in terms of the distance from the set of, respectively, the separable and the classical-quantum states~\cite{MPSVW2010,CBLA2014,CBRLA2014}. Alternatively, one might consider the approach defining quantum correlations as the corresponding entanglement that is necessarily created between system and apparatus during local measurements~\cite{PA2012}.

{\em Acknowledgments} - The authors acknowledge financial support from the Italian
Ministry of Scientific and Technological Research under the PRIN 2010/2011 Research Fund, and from the EU FP7 STREP Projects iQIT, Grant Agreement No. 270843, and EQuaM, Grant Agreement No. 323714. SMG acknowledges financial support from the Austrian Science Foundation, Grant FWF-P23627-N16.

\end{document}